\newif\iffigs\figstrue
\documentclass[paper, 12pt, letterpaper, epsf]{JHEP}
\input{epsf.tex}
\usepackage{graphics}
\def\Bbb{\bf}
\def\C{{\Bbb C}}

\def\Z{{\Bbb Z}}

\def\bearray{\begin{eqnarray}}
\def\eearray{\end{eqnarray}}
\def\bearraynn{\begin{eqnarray*}}
\def\eearraynn{\end{eqnarray*}}
\def\bfig{\begin{figure}}
\def\efig{\end{figure}}

\def\opeq#1{\advance\lineskip#1 \advance\baselineskip#1
        \advance\lineskiplimit#1}

\def\ker{{\rm ker}}

\def\im{{\rm im}}

\newtheorem{Proposition}{Proposition}[section]

\newtheorem{Theorem}{Theorem}[section]
\newtheorem{Lemma}{Lemma}[section]
\newtheorem{Corrolary}{Corrolary}[section]

\newcommand{\be}{\begin{equation}}
\newcommand{\ee}{\end{equation}}
\newcommand{\bea}{\begin{eqnarray}}
\newcommand{\eea}{\end{eqnarray}}

\newcommand{\bp}{\begin{Proposition}}
\newcommand{\ep}{\end{Proposition}}
\newcommand{\bt}{\begin{Theorem}}
\newcommand{\et}{\end{Theorem}}
\newcommand{\bl}{\begin{Lemma}}
\newcommand{\el}{\end{Lemma}}
\newcommand{\bc}{\begin{Corrolary}}
\newcommand{\ec}{\end{Corrolary}}
\newcommand{\nn}{\nonumber}

\font\mybb=msbm10 at 12pt

\def\bb#1{\hbox{\mybb#1}}

\def\P{\bb{P}}
\def\Z {\bb{Z}}

\def\id{\protect{{1 \kern-.28em {\rm l}}}}

\def\boldzeta{\mbox{\boldmath $\zeta$}}

\def \ov {\over }
\def\bea{\begin{eqnarray}}
\def\eea{\end{eqnarray}}

\def\k{{\bf k}}

\def\be{\begin{equation}}
\def\ee{\end{equation}}
\def\ba{\begin{eqnarray}}
\def\ea{\end{eqnarray}}

\def\1{{{(1)}}}\def\2{{{(2)}}}\def\3{{{(3)}}}

\font\mybb=msbm10 at 12pt

\def\bb#1{\hbox{\mybb#1}}

\def\Z {\bb{Z}}

\def\id{\protect{{1 \kern-.28em {\rm l}}}}

\usepackage{graphics}

\title{Holomorphic potentials for graded D-branes}

\author{C.~I.~Lazaroiu
\\C.~N.~Yang Institute for Theoretical Physics\\
SUNY at Stony BrookNY11794-3840,
U.S.A.\\calin@insti.physics.sunysb.edu}

\author{R.~Roiban
\\Department of Physics, UCSB\\
Santa Barbara, CA 93106,
U.S.A.\\radu@vulcan.physics.ucsb.edu}

\abstract{We discuss gauge-fixing, propagators and 
effective potentials for 
topological A-brane composites in Calabi-Yau compactifications. 
This allows for the construction of a holomorphic 
potential describing the low-energy dynamics of such systems,
which generalizes the superpotentials known from the ungraded case. 
Upon using results of homotopy
algebra, we show that the string field and low energy descriptions of 
the moduli space 
agree, and that the deformations of such backgrounds are described by 
a certain extended version of `off-shell Massey products' associated with 
flat graded superbundles. 
As examples, we consider a class of graded 
D-brane pairs of unit relative grade. 
Upon computing the holomorphic potential, we 
study their moduli space of composites.
In particular, we give a general proof that such pairs can form acyclic
condensates, and, for a particular case, 
show that another branch of their moduli space describes 
condensation of a two-form.
}

\preprint{YITP-SB 01-61}

\begin{document}

\tableofcontents

\pagebreak
\vskip .6in

\section{Introduction}

Calabi-Yau compactifications of type II strings in the presence of D-branes
form an interesting class of superstring vacua 
in four dimensions, with rich potential applications 
for string phenomenology. 
Such compactifications
have recently attracted a great deal of attention \cite{Douglas_Kontsevich, 
Aspinwall, LSM, Douglas_quintic, pi_stab, Kachru, LSM, Douglas_Beilinson, 
Douglas_Diaconescu, Douglas_Aspinwall}. While compactifications 
in the presence of one brane are at least conceptually well-understood, 
the situation is rather different for backgrounds containing more general 
D-brane configurations, whose systematic study has begun only recently. 
One of the central problems in the subject is the issue of D-brane 
composites, i.e. bound states of various D-branes resulting upon 
condensation of spacetime fields associated with boundary condition changing 
sectors \cite{Douglas_Kontsevich, Aspinwall, Oz_superconn, Oz_triples}. 
The basic process of this type (namely tachyon condensation)
is known to lead to a wealth of D-brane composites, whose 
systematic analysis is rather involved.

Perhaps the most powerful approach to this subject has been 
proposed in \cite{Douglas_Kontsevich}. The 
strategy is to separate the problem into a `topological step'
(described by the associated twisted models
\cite{Witten_nlsm, Witten_mirror, Witten_CS}), 
which allows one to classify all composites 
resulting from condensation of space-time fields associated with 
open string chiral primaries, and a 
condition \cite{Douglas_Kontsevich, Douglas_Aspinwall}, 
whose role is to identify those composites 
which are stable under decay. 

In view of this description, a better understanding of  
D-brane condensates requires a detailed study of their topological 
avatars. While the spectrum of topological composites  is by now relatively 
well-established (being described by objects of certain categories
naturally associated with the closed string background), rather little 
is currently known about another basic aspect, namely their {\em moduli space}.

The purpose of this paper is to initiate the study of such moduli 
spaces. The approach we propose will be based on a particularly 
explicit formulation of D-brane dynamics which describes the formation 
of such composites within the framework of topological string field
theory \cite{com1, com3, Diaconescu, sc, bv}. This description 
results from the basic observation of \cite{Seidel, Zaslow_Polishchuk} that 
topological D-branes of a Calabi-Yau compactification are {\em graded} 
objects.
The models of  \cite{Diaconescu, sc, bv} are based 
on a certain extension of the topological field theories of open A/B 
strings \cite{Witten_CS}, which is devised to take into account the novel data 
provided by the D-brane grades. This can be 
formulated as a `graded Chern-Simons field theory', 
a version of Chern-Simons theory based on a graded superbundle\footnote{
We stress that this does {\em not} coincide with the super-Chern-Simons field
theory considered in \cite{Vafa_cs}.}.

A preliminary study of these models was carried out in 
\cite{com3, Diaconescu, sc}, which gave arguments relating their moduli 
spaces to certain enhanced triangulated categories naturally associated 
with the problem. It was also argued in  \cite{sc} that the moduli space 
of such theories can be viewed as a certain `extended' version of the
moduli space of ungraded D-branes. In this description, `nonstandard' 
directions in the extended moduli space correspond to condensation processes,
and nonstandard moduli points are associated with topological brane 
composites. Hence one may study such points with usual 
field-theoretic tools. Moreover, it was showed in
\cite{bv} that graded Chern-Simons theories are consistent as classical
BV systems, and thus form a good starting point for a quantum analysis. 
 
From a mathematical perspective, the moduli spaces discussed
in \cite{sc, bv} correspond to a deformation problem, whose local description
is of Maurer-Cartan type. 
Since the resulting deformations may be obstructed, 
a detailed analysis of the moduli space
requires a systematic study of obstructions. 

In fact, obstructed deformations 
are also common in ungraded D-brane systems. A method of dealing 
with obstructions (as they apply to the ungraded case) 
was proposed in \cite{superpot}, where it was shown that the correct moduli 
space can be described in terms of a certain potential for the massless
modes \footnote{The idea of such a potential 
goes back to the work of \cite{Witten_CS}, and is also implicit 
in \cite{Kontsevich_Felder}.}. 
This potential, which is extremely 
natural from a string field-theoretic point of view, 
coincides with the D-brane superpotentials of \cite{Douglas_quintic} 
(see also \cite{Kachru}). 
Moreover, the potential 
is intimately related to certain constructions of modern deformation theory 
\cite{Kontsevich_Felder, Merkulov_infty, Merkulov_defs, 
Barannikov_Kontsevich} (see also \cite{Barannikov}), 
which were used to establish the main result of \cite{superpot}. 
One advantage of this approach is that it allows for a description 
of the moduli space which does not involve differential equations
--and thus is often more effective from a computational point of view.
As explained for example in \cite{Merkulov_infty, Merkulov_defs}, this 
is intimately connected with standard constructions
of Kuranishi theory \cite{Kuranishi}.

Since the definition of the potential given in \cite{superpot} 
is purely field-theoretic in nature, one expects that it 
extends to the graded theories of \cite{sc, bv, Diaconescu}. 
In this paper, we show that this is indeed
the case. In particular, we shall build a holomorphic potential 
which is a natural extension of the brane superpotentials of \cite{superpot} 
to the case of graded D-branes.  
Moreover, the arguments of \cite{superpot} can be adapted 
to this situation in order to show that the holomorphic 
potential we construct provides an equivalent description of the moduli 
space. This allows us to 
determine the moduli space of topological composites for a particular 
class of graded D-brane pairs of unit relative grade.

The present paper is organized as follows. 
In Section 2, we discuss some basic aspects of the theories under 
study. The models we consider describe arbitrary collections of 
topological D-branes wrapping a given special Lagrangian 
cycle of a Calabi-Yau threefold compactification, and provide a description
of chiral primary dynamics\footnote{We consider the large radius limit 
only, in order to avoid instanton corrections to the string field 
action.}. After reviewing the 
origin and structure 
of such theories, we discuss their gauge symmetries and moduli spaces.
We also construct a certain conjugation operator and 
characterize the associated zero modes in terms
of Hodge theory. 
Section 3 considers the example  of D-brane pairs with unit relative grade. 
Upon using the methods of Section 2, we analyze harmonic modes of 
such a system. This leads to a completely general 
discussion of acyclic composites, which extends certain results
of \cite{bv}. We also give a preliminary analysis of the relevant moduli 
spaces. 
In Section 4, we 
consider the partition function of our models and give a physical 
definition of a holomorphic potential for virtual deformations (a.k.a. 
massless, or harmonic modes). Upon using a straightforward extension of 
the gauge-fixing condition employed in \cite{superpot}, we discuss the 
tree-level expansion of the potential around a classical solution and 
the algebraic structure of scattering products.
Applying results of modern deformation theory, we 
show that the moduli space can be described locally as a quotient of the 
critical set of the potential through the action of certain symmetries. 
This generalizes the results of \cite{superpot} to the case of graded 
D-branes. The results of this section 
assume consistency of our gauge-fixing procedure.
A general proof of this statement, which requires the 
Batalin-Vilkovisky formalism, turns out to be quite technical and 
will be given in a companion paper \cite{gf}. The main results of that 
analysis are summarized in Subsection 4.2.
Section 5 considers the application of our methods to 
graded D-brane pairs (of unit relative grade) on a three-torus. 
For the singly-wrapped case, 
we compute the holomorphic potential and its symmetry group,
and give a local description of the moduli space.
We also make a few observation about the multiply-wrapped case. 
Section 6 presents our conclusions and a few directions for further 
research. Appendix A contains some technical details, while Appendix B gives
an alternate (but equivalent) construction of the tree-level approximation 
to the holomorphic potential.

\section{Structure of the string field theory}

Consider a special Lagrangian 3-cycle $L$ of a Calabi-Yau threefold $X$.
Throughout this paper, we assume that $L$ is 
connected. 
We shall be interested in systems of graded topological D-branes 
(of different grades) wrapping this cycle. As argued in \cite{sc}, such 
systems are described (in the large radius limit) 
by a string field theory which is a graded form 
of Chern-Simons field theory. To formulate this, 
we first review the 
`BPS grade' of \cite{pi_stab, 
Douglas_Kontsevich, Aspinwall, Douglas_Aspinwall} and its relation 
with a choice of orientation of the cycle \cite{sc}. 

\subsection{The BPS grade}

Let $\Omega$ be the holomorphic 3-form of $X$, normalized such that
$\frac{\omega^3}{3!}=\frac{i}{8}\Omega\wedge {\overline \Omega}$.
In this case, $\Omega$ is determined up to a phase, which we fix 
for what follows. 
Recall that a Lagrangian cycle $L$ is {\em special Lagrangian} 
(this description follows from Proposition 2.11 of \cite{Joyce}) if there 
exists a complex number $\lambda$ of unit modulus such that:
\be
Im(\lambda \Omega)|_{L}=0~~.
\ee 
This condition determines $\lambda$ up to a sign ambiguity 
($\lambda\rightarrow -\lambda$), so that only the complex quantity 
$\lambda^2$ is naturally defined by the (unoriented) cycle $L$. 
Given a choice for $\lambda$, the real 3-form $\lambda\Omega|_L$
is nondegenerate and thus induces an orientation ${\cal O}_\lambda$
of $L$. When endowed with this orientation, $L$ is a calibrated submanifold 
of $X$ with respect to the calibration given by $Re(\lambda\Omega)$. 
In particular, one has:
\be
Re(\lambda\Omega)|_L=\lambda \Omega|_L=vol_{L, {\cal O}_\lambda}~~,
\ee
where $vol_{L, {\cal O}_\lambda}$ is the volume form on $L$ induced by the 
Calabi-Yau metric of $X$ and the orientation ${\cal O}_\lambda$. 
We have $\int_{L, {\cal O}_L}\Omega=\frac{\beta}{\lambda}$, 
where $\beta=vol(L)=|\int_L\Omega|>0$ 
is a positive number. Hence:
\be
\lambda^{-1}=\frac{\int_{L,{\cal O}_\lambda}\Omega}{|\int_L\Omega|}
\Leftrightarrow \lambda_{\cal O}^{-1}=
\frac{\int_{L,{\cal O}}\Omega}{|\int_L\Omega|}~~,
\ee
which shows how an orientation ${\cal O}$ of $L$ determines $\lambda$. 

Following \cite{Aspinwall, Douglas_Aspinwall}, 
we define the {\em BPS grade} of the {\em oriented} cycle $(L, {\cal O})$ 
via: 
\be
\label{BPS_grade}
\phi_{L, {\cal O}}=
\frac{1}{\pi}{\rm arg}\int_{L, {\cal O}}\Omega=
\frac{1}{\pi}{\rm Im} \log \int_{L, {\cal O}}\Omega~~.
\ee
Note that $\phi_{L, {\cal O}}$ only depends on the homology class 
$[L, {\cal O}]$ of 
the oriented cycle $(L, {\cal O})$ (the pushforward of its fundamental 
class through the inclusion map).  
With this definition, we have:
\be
\lambda_{\cal O}=e^{-i\pi \phi_{L, {\cal O}}}
\Rightarrow \lambda^2=e^{-2i\pi \phi_{L, {\cal O}}}~~. 
\ee
It is clear that $\phi_{L, {\cal O}}$ is determined by $(L, {\cal O})$ 
only up to a shift by $2$. Moreover:
\be
\label{opposite}
\phi_{L, -{\cal O}}=\phi_{L, {\cal O}}+1~~(mod~2)~~.
\ee
Note that there exists a `fundamental' choice of grading, namely 
$\phi\in [0,1)$. This induces the `fundamental orientation' ${\cal O}_0$ 
for which $Im (\lambda^{-1}_{{\cal O}_0})>0$.

It can be argued in various ways that a consistent description of D-brane 
systems forbids one from restricting $\phi$ to a fundamental domain (of length 
two). 
The basic picture is as follows. Following the approach of \cite{top, com1}, 
we are given a closed conformal field theory (parameterized by the 
complex and Kahler structure of $X$), whose moduli space ${\cal Q}$ 
can be described 
as a product between the complex structure moduli space of $X$ and that 
of its mirror. For any point in this moduli space (i.e. for a fixed closed 
CFT background), we are interested in the category of all D-branes compatible 
with this bulk CFT; in the language of \cite{top, com1}, 
this is the category of open-closed extensions of the bulk theory.
Invariance of our physical description requires that this category ${\cal C}$ 
be well-defined at every point in ${\cal Q}$. 
In general, a given D-brane cannot be deformed to an object which is 
stable throughout this moduli space; moreover, monodromies around the 
discriminant loci will act nontrivially on the category ${\cal C}$. The 
requirement of a well-defined theory implies that the monodromy group be 
represented through `autoequivalences' (in an appropriate sense) of 
${\cal C}$. This condition must hold both at the topological level (i.e. 
if ${\cal C}$ consists of all topological D-branes compatible with the 
bulk theory) and at the level of stable BPS branes. Restricting to the 
topological D-brane category, it is clear that such monodromies will 
shift the grade of various objects outside of any fundamental domain, which 
is why one must consider all branches of (\ref{BPS_grade})\footnote{A 
monodromy action typically transforms a D-brane described by a 
(cycle, bundle) pair into a D-brane composite (obtained from a collection 
of (cycle, bundle) pairs by condensation of fields supported at their 
intersections. Such composites are related to Fukaya's category 
\cite{Fukaya, Fukaya2}. On the other hand, some other composite will generally 
be transformed into the original (cycle, bundle) pair (with a shifted grade) 
upon the same monodromy transformation.}.
The monodromy action should be viewed as a group of discrete gauge-invariances 
of the topological D-brane theory.  

Using such monodromy actions (for example, monodromy around a large complex 
structure point of $X$, with the Kahler parameters kept close to large radius),
one can generally 
produce objects of ${\cal C}$ based on the cycle $L$, but whose 
grades are shifted by $2n$ for any $n\in \Z$; we shall call such objects the 
`shifts' of $L$. As argued in \cite{com1, com3}, 
the full category of topological D-branes must in fact be enlarged to the 
collection of all topological brane composites, which result by considering 
all condensation processes of spacetime-fields 
associated with topological boundary condition changing operators.  
A monodromy-invariant description requires that we consider 
condensates between an object and its shifts. In this paper, 
we are interested in the sector of 
topological string field theory which describes the 
formation of such condensates\footnote{In the mirror picture, this 
is the sector describing a B-type brane based on a coherent sheaf ${\cal E}$
(or a complex of such)  
and all of its shifts. Note that we do not pass to the derived category. 
The existence of graded objects and of shift functors is a {\em prerequisite} 
of the derived category description. In fact, the arguments of 
\cite{Douglas_Kontsevich, Aspinwall} are based on this assumption. 
For certain problems, the derived category language is not the 
most advantageous. The 
problem of deformations, which is the main focus of this paper, is one such 
example.}. 

Given a special Lagrangian cycle $L$, let us thus consider the collection of 
all its shifts $L[2n]$ (with $n\in \Z$). Since changing the orientation of
$L$ is related (modulo a change in GSO projections) 
to passing from a brane to its antibrane, and in view of relation 
(\ref{opposite}), one must in fact also consider shifts $L[2n+1]$ by odd 
integers. Hence a complete description of D-branes wrapping $L$ must 
consider all integral shifts of $L$. For any such brane $L[n]$, 
one also specifies 
a background given by a flat connection $A_n$ in a complex vector bundle 
$E_n$ over $L$. Since we work with the topological model, there is no need 
to impose (anti) hermicity conditions on $A_n$ (we shall allow 
the string field action to be complex
\footnote{This 
is quite standard for the (ungraded) 
topological open B-model, whose action is complex 
as well.}). Given this data, it was argued in \cite{sc} that the topological 
string field theory of the D-brane collection $(L[n], A_n)$ 
is a `graded Chern-Simons model', which we now explain.

\subsection{Ingredients of the topological model} 

Given a collection of graded branes (of different grades $n$) wrapping $L$, 
we form the total bundle ${\bf E}=\oplus_{n}{E_n}$, endowed with the 
$\Z$-grading induced by $n$. 
As argued in \cite{sc}, the presence of 
D-brane grades which differ by integers 
shifts the assignment of worldsheet $U(1)$ charge 
in the various boundary condition changing sectors.
The result is that the charge of states of strings stretching from $E_m$ 
to $E_n$ is shifted by $n-m$. Since such states localize \cite{Witten_CS} 
on differential 
forms on $L$, valued in $Hom(E_m, E_n)$, the worldsheet charge induces a 
grading $|.|$ 
on the space of $End({\bf E})$-valued forms:
\be
|u|=rk u + \Delta(u)~~,
\ee
where $\Delta(u)=n-m$ if $u\in \Omega^*(L, Hom(E_m,E_n))$. 
In geometric terms, we are interested in sections $u$ 
of the bundle: 
\be
{\cal V}=\Lambda^*(T^*L)\otimes End({\bf E})~~,~~
\ee
endowed with the grading ${\cal V}=\oplus_{t}{{\cal V}^t}$, where:
\be
{\cal V}^t=\oplus_{\tiny \begin{array}{c}k, m, n\\k+n-m=t\end{array}}
{\Lambda^k(T^*L)\otimes Hom(E_m, E_n)}~~.
\ee
The space ${\cal H}=\Gamma({\cal V})$ of such sections is the 
{\em total boundary space} of \cite{sc}, and can be interpreted as the 
collection 
of all open string states of the system. 
It is endowed with the grading ${\cal H}^k=\Gamma({\cal V}^k)$.

The second ingredient of \cite{sc} is the so-called 
{\em total boundary product}, which is defined through:
\be
u\bullet v= (-1)^{\Delta(u)rk v}u\wedge v~~,
\ee
where the wedge product on the right hand side includes composition 
of morphisms in $End({\bf E})$. As discussed in \cite{sc} that this 
product is associative (albeit not commutative, in general). It also 
admits the identity endomorphism $1$ of ${\bf E}$ 
as a neutral element:
\be
1\bullet u=u\bullet 1=u~~.
\ee
Moreover, the product is compatible with the grading on ${\cal H}$:
\be
|u\bullet v|=|u|+|v|~~
\ee
(note that $|1|=0$), and thus endows this space with a structure of graded 
associative algebra. 

The third ingredient arises by noticing that
${\bf E}$ is endowed with a flat structure, the direct sum of flat 
structures on the bundles $E_n$. This can be described by the direct 
sum connection $A^{(0)}=\oplus_n{A_n}$.  The flat connection $A^{(0)}$
determines a nilpotent differential $d$ on ${\cal H}$, the de Rham 
differential coupled to the connection induced by $A^{(0)}$ on $End({\bf E})$. 
This operator acts as a degree one 
derivation of the boundary product:
\be
|du|=|u|+1~~,~~d(u\bullet v)=(du)\bullet v+(-1)^{|u|}u\bullet (dv)~~.
\ee
Endowed with the product $\bullet$ and this differential, ${\cal H}$ becomes 
a {\em differential graded associative algebra} (dGA). 

The final ingredient of \cite{sc} is a bilinear form on ${\cal H}$ induced 
by the graded trace  on the bundle $End({\bf E})$. The latter is 
defined through:
\be
str(u)=\sum_{n}{(-1)^n tr(u_{nn})}~~,~~{\rm for}~~u=\oplus_{m,n}{u_{mn}}~~,
\ee
with $u_{mn}\in \Omega^*(L, Hom(E_m, E_n))$. This associates a 
form with complex coefficients to every $End({\bf E})$-valued form on $L$.
The bilinear form:
\be
\label{bf}
\langle u, v \rangle:=\int_{L}{str(u\bullet v)}~~ 
\ee
is non-degenerate and has the properties:
\bea
&&\langle u, v\rangle =(-1)^{|u||v|}\langle v, u \rangle~~\nn\\
&&\langle du, v\rangle +(-1)^{|u|}\langle u, dv\rangle=0~~\\
&&\langle u\bullet v, w\rangle =\langle u, v\bullet w\rangle~~.\nn 
\eea
In words, it is a graded-symmetric, invariant bilinear form on the differential
graded algebra $({\cal H}, d, \bullet)$. In (\ref{bf}) and in all 
other integrals over $L$, we assume that the special Lagrangian cycle has been 
endowed with its `fundamental' orientation ${|cal O}_0$ 
(see \cite{sc} and Subsection 2.1).

\subsection{The action}

The string field theory of \cite{sc,bv} is described by the action: 
\be
\label{action}
S(\phi)=Re\int_{L}{str\left[\frac{1}{2}\phi\bullet d\phi+\frac{1}{3}
    \phi\bullet\phi\bullet\phi\right]}=
\frac{1}{2}\langle \phi, d\phi\rangle +\frac{1}{3}\langle
\phi,\phi\bullet \phi\rangle~~~~,  
\ee which is defined on the degree one
component 
\be
{\cal H}^1=\{\phi\in {\cal H}||\phi|=1\}=
\Gamma(\oplus_{k+n-m=1}{\Lambda^k(T^*L)\otimes Hom(E_m,E_n)})
\ee 
of the total
boundary space. This action defines a `graded Chern-Simons field theory', 
which is related to, but {\em not} 
identical with\footnote{The major difference is 
that the theories of \cite{supergroupCS, Vafa_cs} contain 
only physical fields of rank one, while our models 
will typically contain physical fields of all ranks. It should 
be noted that the proposal of \cite{Vafa_cs} would require condensation 
of ghosts and/or antifields, which does not seem to be a physically 
meaningful process.}, the super-Chern-Simons 
field theory considered in \cite{supergroupCS, Vafa_cs}. The 
physics described by
(\ref{action}) is considerably more complicated than that of usual or 
super-Chern-Simons theories.

\subsection{Gauge symmetries}

The theory (\ref{action}) is invariant with respect to a gauge group 
which can be described as follows. Since the boundary product 
is compatible with the degree $|.|$, it follows that 
the subspace 
${\cal H}^0=
\Gamma(\oplus_{k+n-m=0}{\Lambda^k(T^*L)\otimes Hom(E_m,E_n)})$ of 
charge zero elements of ${\cal H}$ forms a subalgebra of the total 
boundary algebra $({\cal H}, \bullet)$. Since $|1|=0$, this subalgebra 
has a unit. It follows that the set: 
\be
{\cal G}=\{g\in {\cal H}^0|{\rm~exists~}~g^{-1}\in {\cal H}^0~{\rm~such~that~}
~g\bullet g^{-1}=g^{-1}\bullet g=1\}~~
\ee
of invertible 
elements of $({\cal H}^0,\bullet)$ forms a group with respect to the 
boundary multiplication.
Its adjoint action: 
\be
\label{Adjoint}
u\rightarrow Ad_g(u):=g\bullet u\bullet 
g^{-1}~~,~~{\rm~for~}~~u\in {\cal H}~~,~~g\in {\cal G}
\ee
on the total boundary space preserves  the worldsheet degree 
$|.|$, and in particular induces an action on the 
subspace ${\cal H}^1$ of degree one states. 

If $g$ is close to the identity, then one can use the exponential parameterization:
\be
\label{exp}
g=e^\alpha_\bullet :=\sum_{k\geq 0}{\frac{1}{k!}\alpha^{\bullet k}}~~,
\ee
where $\alpha^{\bullet k}$ stands for $k$-th iteration of 
the $\bullet$-product of $\alpha$ with itself (and we define 
$\alpha^{\bullet 0}:=1$). In particular, the Lie algebra 
of ${\cal G}$ can be described as follows. For any two elements 
$u, v$ of ${\cal H}$, we define their {\em graded commutator}
by:
\be
[u,v]_\bullet:=u\bullet v -(-1)^{|u||v|}v\bullet u~~.
\ee
This bracket is graded antisymmetric:
\be
[u,v]_\bullet =-(-1)^{|u||v|}[v,u]_\bullet
\ee
and satisfies the graded Jacobi identity:
\be
[[u,v]_\bullet,w]_\bullet+(-1)^{|u|(|v|+|w|)}[[v,w]_\bullet, u]_\bullet+ 
(-1)^{|w|(|u|+|v|)}[[w,u]_\bullet, v]_\bullet=0~~,
\ee
as well as the relation:
\be
d[u,v]_\bullet=[du,v]_\bullet +(-1)^{|u|}[u, dv]_\bullet~~,
\ee
thereby making ${\cal H}$ into a differential graded Lie algebra (dGLA). 
The subspace ${\cal H}^0$ of degree zero elements is closed under the bracket, 
and forms a {\em usual} Lie algebra with respect to the induced operation, 
which on degree zero elements coincides with the standard commutator:
\be
[\alpha,\beta]_\bullet =\alpha\bullet \beta-\beta\bullet \alpha~~{\rm~~for~~}~~
\alpha,\beta\in {\cal H}^0~~.
\ee
It is clear that $({\cal H}^0,[.,]_\bullet)$ coincides with the Lie algebra 
of the gauge group ${\cal G}$. Differentiating (\ref{Adjoint}) 
shows that this algebra acts on ${\cal H}$ through its adjoint representation:
\be
u\rightarrow ad_\alpha(u):=[\alpha, u]_\bullet~~.
\ee

By analogy with usual Chern-Simons theory, we consider the gauge 
transformations:
\be
\label{Gauge}
\phi\rightarrow \phi^g=g\bullet \phi \bullet g^{-1}+g\bullet dg^{-1}~~.
\ee
Upon using the 
derivation property of $d$, the identity $g\bullet g^{-1}=g^{-1}\bullet g=1$
implies:
\be
dg^{-1}=-g^{-1}\bullet (dg)\bullet g^{-1}~~.
\ee
Combining this identity with the invariance properties of the bilinear form, 
one can check that the action 
(\ref{action}) transforms as follows under (\ref{Gauge}):
\be
S(\phi)\rightarrow S(\phi^g)=S(\phi)+\Delta(g)~~,
\ee
where:
\be
\Delta(g)=\frac{1}{6}\int_{L}{
str(g^{-1}\bullet dg\bullet g^{-1}\bullet dg \bullet g^{-1}\bullet dg)}~~.
\ee

For infinitesimal $\alpha$, the gauge 
transformations (\ref{Gauge}) become:
\be
\label{gauge}
\phi\rightarrow\phi+\delta_\alpha\phi~~, 
\ee 
with $\delta_\alpha\phi=-d\alpha-[\phi,\alpha]_\bullet$, and one can directly 
check the relation:
\be
\delta_\alpha\delta_\beta\phi-\delta_\beta\delta_\alpha\phi=\delta_{
  [\alpha,\beta]_\bullet}\phi~~,
\ee
i.e. the gauge algebra closes off shell. For an infinitesimal 
transformation (\ref{gauge}), one has $g^{-1}\bullet dg=d\alpha +O(\alpha^2)$
and thus the quantity $\Delta(g)$ vanishes to third order in $\alpha$:
\be
\Delta(g)=\frac{1}{6}\int_{L}{str(d\alpha\bullet d\alpha\bullet d\alpha)}
+O(\alpha^4)
=\frac{1}{6}\int_{L}{dstr(\alpha\bullet d\alpha\bullet d\alpha)}+O(\alpha^4)=
O(\alpha^4)~~.
\ee
This implies:
\be
\frac{d}{dt}S(\phi^{e^{t\alpha}_\bullet})=0
\Rightarrow S(\phi^{e^{t\alpha}_\bullet})=S(\phi)~~{\rm~for~}~~
\alpha\in {\cal H}^0~~.
\ee
Taking $t=1$ shows that the action \ref{action} is invariant\footnote{In the 
case of usual Chern-Simons theory, one uses a formulation in terms of 
principal bundles and shows that the action transforms 
by integer shifts (in appropriate units) under large gauge transformations, 
and thus the path integral is gauge-invariant in this general sense. 
It is likely that a similar result holds true for our theories. Instead of 
attempting a proof, we shall be pragmatic and restrict to small 
gauge transformations. This will suffice for the 
perturbative analysis of Section 4.}  
under {\em small gauge transformations}, which we define as those gauge 
transformations which can be written in the exponential form (\ref{exp}).

We finally note that the adjoint action (\ref{Gauge}) 
of a `small' element $g$ can be written as:
\be
Ad_{e^\alpha_\bullet}=e^{ad_\alpha}~~,
\ee
where the right hand side is the formal exponential of $ad_\alpha$ viewed as a linear 
operator in the vector space ${\cal H}$. 
Moreover, the gauge group action (\ref{Gauge}) takes the form:
\be
\phi\rightarrow \phi^{e^\alpha_\bullet}=
e^{ad_\alpha}\phi-\frac{e^{ad_\alpha}-1}{ad_\alpha}d\alpha~~,
\ee
where the fraction in the last term is formally defined by its power 
series expansion (better, by functional calculus). This recovers the 
formulation used in \cite{superpot}. 

\paragraph{The ungraded case}

Since our description of the gauge group may seem unfamiliar, let us 
consider what it becomes in the ungraded case. This corresponds to 
${\bf E}=E_0$, i.e. a single ungraded D-brane wrapping $L$. 
In this situation, (\ref{action}) reduces to the usual Chern-Simons theory 
coupled to the bundle $E_0$ (and expanded around the background 
flat connection $A_0$). The degree zero component of the total boundary algebra
is ${\cal H}^0=\Gamma(End(E_0))$, the space of endomorphisms of the bundle $E_0$; 
this is endowed with the multiplication given by usual fiberwise composition 
of morphisms:
\be
\alpha\bullet \beta=\alpha\circ \beta~~.
\ee
The units of this algebra form the standard gauge 
group ${\cal G}=\Gamma(Aut(E_0))$ of automorphisms 
of $E_0$.  
Our presentation of ${\cal G}$ in the graded case 
is the generalization of this description.

\subsection{The classical moduli space and its interpretation}

The critical points of (\ref{action}) are solutions to the equation:
\be
\label{mc}
\frac{\delta S}{\delta \phi}=0\Longleftrightarrow 
d\phi+\frac{1}{2}[\phi,\phi]_\bullet=0~~,
\ee
with $\phi$ an element of ${\cal H}^1$ (note that 
$\frac{1}{2}[\phi,\phi]_\bullet =\phi\bullet \phi$ for a degree one element 
$\phi$). The equation of motion (\ref{mc}) is invariant under the 
gauge group action (\ref{Gauge}) (with $g$ an arbitrary gauge transformation,
small or large).
The moduli space ${\cal M}$ results upon dividing the space of solutions 
through this gauge group action. The Maurer-Cartan 
condition (\ref{mc}) describes deformations of the reference 
connection $A^{(0)}$ into a `flat superconnection $A$ of total degree one'
(in the sense of \cite{Bismut_Lott}). 
Hence ${\cal M}$ is the moduli 
space of such superconnections, defined on the graded bundle ${\bf E}$. 
The original background $A^{(0)}$ corresponds to a 
`diagonal' superconnection, constructed as a direct sum of flat connections 
on the bundles $E_n$.

The adjoint and gauge-group actions (\ref{Adjoint},\ref{Gauge}) 
preserve the degree $|.|$. The total grading of the 
bundle ${\cal V}$ is a gauge-invariant concept, and the collection 
${\cal S}$ of flat degree one superconnections is well-defined.
The gauge-group action (\ref{Gauge}) preserves ${\cal S}$, and 
the moduli space ${\cal M}$ can be defined geometrically as
the quotient ${\cal M}={\cal S}/{\cal G}$. 
A reference point is only 
necessary when writing the
Maurer-Cartan equation (\ref{mc}). In physical terms, a reference 
superconnection appears because we use a background-dependent formulation 
of the string field theory. 

In general, ${\cal M}$ is a rather complicated
object, which depends on the topology of $L$ and on the 
structure of the graded superbundle ${\bf E}$. It is clear that this moduli 
space can have singularities or fail to be compact, and that a global study 
requires some careful analysis in the manner familiar from the usual theory of 
flat connections. 

Following \cite{com1, com3} and \cite{sc}, we recall the D-brane 
interpretation of ${\cal M}$. 
As argued in those papers, an off-diagonal background 
corresponds to condensation of the spacetime fields associated with the 
combination of boundary condition changing operators described by 
the string field $\phi$. This process leads to the formation 
of D-brane composites, thus altering the brane interpretation of the 
background. 
The main observation is that an off-diagonal background violates the original 
decomposition of the total boundary space into the sectors 
$\Gamma(\Lambda^*(T^*L)\otimes Hom(E_m, E_n))$, and thus alters the 
D-brane content 
of the theory. When expanded around the new background, the string field
action  has the same form (\ref{action}) (up to addition of an irrelevant constant), 
but with a shifted differential:
\be
d\rightarrow d_\phi~~,~~d_\phi u=du +[\phi,u]_\bullet~~.
\ee
As explained in \cite{com1, com3}, the 
new D-brane content can be identified by studying certain decomposition 
properties of the shifted differential, together with the boundary product and 
bilinear form. This can be 
discussed systematically in the language of category theory \cite{com1, com3}.

\subsection{Virtual dimensions and obstructions}

The linearization\footnote{This is obtained by assuming that {\em both} $\phi$ and $\alpha$ 
are small, and keeping only the first order contributions.}
of (\ref{mc}) and (\ref{Gauge}) is specified by:
\be
\label{linearized}
d\phi=0~~,~~\phi\equiv \phi -d\alpha~~. 
\ee
Thus first order deformations are in one to one 
correspondence with elements of 
the cohomology group $H_d^1({\cal H}):=\ker(d:{\cal H}^1\rightarrow {\cal H}^2)/
\im(d:{\cal H}^0\rightarrow {\cal H}^1)$. This gives the virtual dimension:
\be
\label{vdim}
vdim_A{\cal M}:=h_1~~,
\ee
where $h_k$ denotes the dimension of $H^k_{d_A}({\cal H})$. The 
approximation (\ref{linearized}) need of course not suffice, and  
typically some infinitesimal deformations will be obstructed.
Such obstructions lift some directions in $H^1_d({\cal H})$, 
leading to a moduli space of dimension smaller than (\ref{vdim}).

As in \cite{superpot}, our approach to obstructions will be to
construct a 
function $W$ (the holomorphic potential of Section 4), 
which is defined on the space 
$H^1_{d_A}({\cal H})$ of {\em virtual} deformations 
(the `virtual tangent space' to ${\cal M}$ at $A$) and 
whose critical set gives a {\em local} description of 
the true moduli space 
after dividing out through some effective symmetries. 
From this perspective, the potential 
$W$ is a tool for 
dealing with obstructions in a systematic manner. 

\subsection{Harmonic analysis of linearized zero modes}

It is convenient to describe the space of virtual 
deformations through harmonic analysis. 
In this subsection, we develop the ingredients of such a description, 
which will be useful in later sections. 

\subsubsection{The Hermitian scalar product and conjugation operator}

Let us pick a Riemannian metric $g$ on the three-manifold $L$
and Hermitian metrics on each of the bundles $E_n$, which induce a 
Hermitian metric on the direct sum ${\bf E}=\oplus_{n}E_n$. 
These metrics (which are arbitrarily chosen) 
will be kept fixed for what follows (the physics 
will be independent of all choices).

We first consider the Hermitian
conjugation operator $\dagger$ on sections of the bundle $End({\bf
E})$. This is involutive and antilinear, and inverts the
grading $\Delta$:
\be
(f^\dagger)^\dagger=f~~,~~
(\lambda f)^\dagger={\overline \lambda}f^\dagger~~,
~~\Delta(f^\dagger)=-\Delta(f)~~,
\ee
for $f\in \Gamma(End({\bf E}))$ and $\lambda$ a complex-valued function
on $L$.  We also note the properties:
\be
\label{herm12}
str(f^\dagger)=\overline{str(f)}~~{\rm~and~}~~
(f\circ g)^\dagger=g^\dagger \circ f^\dagger~~.
\ee

On the space $\Omega^*(L)$ of complex-valued forms, we have the {\em complex linear} 
Hodge operator $*$, which is involutive and satisfies:
\be
rk(*\omega)=3-rk\omega~~,
\ee
with $\omega\in \Omega^*(L)$. If $(.,.)$ is the metric induced on
$\Omega^*(L)$ by the Riemannian metric on $L$, then $*$ has
the defining property:
\be
(*\omega) \wedge \eta=(\omega,\eta)vol_g~~,
\ee
where $vol_g$ is the volume form on $L$ induced by $g$.  Since
$(\omega,\eta)=(\eta,\omega)$, this implies the identity:
\be
\label{herm3}
(*\omega)\wedge \eta=(*\eta)\wedge \omega~~.
\ee
Viewing ${\bf E}$ as a usual vector bundle (by forgetting the grading), 
we have a Hermitian scalar product:
\be
h(u,v)=\int_{L}{tr(*u^\dagger\wedge v)}~~,~~u,v\in {\cal H}~~,
\ee
which takes the following form on decomposable elements 
$u=\omega\otimes f$ and $v=\eta\otimes g$:
\be
h(u,v)=\int_{L}{tr(f^\dagger \circ g)(*{\overline \omega}\wedge \eta)}~~.
\ee
Following \cite{superpot}, we look for an antilinear operator $c$ on ${\cal H}$
with the property:
\be
\label{h}
h(u,v)=\langle cu,v\rangle=\int_{L}{str[(cu)\bullet v]}~~.
\ee
Since the bilinear form is non-degenerate, this condition determines 
$c$ uniquely. Upon considering decomposable elements, it is not hard to check
that:
\be
\label{c}
c(\omega\otimes f)=(-1)^{n+\Delta(f)(1+rk \omega)}
(*{\overline \omega})\otimes f^\dagger~~,~~{\rm~for~}~~\omega\in \Omega^*(L)~~
{\rm~and~}~~f\in Hom(E_m,E_n)~~.
\ee
Note the sign factor $(-1)^n$ (which depends on the grade of the image 
of $f$). This is necessary
in order to convert the trace in the definition of $h$ to the graded trace 
appearing in relation (\ref{h}). 
The operator $c$ satisfies:
\be
|cu|=3-|u|~~.
\ee
It is also easy to see\footnote{For this, one notices that 
$rk *{\overline \omega}=3-rk \omega$, 
$\Delta(f)=n-m$ and $f^\dagger\in Hom(E_n, E_m)$.}
that $c$ is involutive (i.e. $c^2=id$).
The Hermitian metric and conjugation operator 
satisfy all requirements of the abstract framework 
discussed in \cite{superpot}. 

\subsubsection{The deformed Laplacian and harmonic modes}

For a linear operator $B$ on ${\cal H}$, we let $B^\dagger$ denote its
Hermitian conjugate with respect to $h$. 
As shown in \cite{superpot}, the properties of $c$ imply:
\be
\label{dder}
d^\dagger u=(-1)^{|u|}cdc u~~{\rm ~and~}~~
\langle d^\dagger u, v\rangle =(-1)^{|u|}\langle u, d^\dagger v\rangle~~.
\ee
We also note that $|d^\dagger u|=|u|-1$. Let us consider the `deformed Laplacian'\footnote{This should not be confused with the partial grading denoted by 
the same letter.}:
\be
\Delta:=d^\dagger d+d d^\dagger~~,
\ee
which is a Hermitian, degree zero, elliptic differential operator of order two 
on ${\cal H}=\Gamma({\cal V})$:
\be
|\Delta u|=|u|~~,~~\Delta^\dagger=\Delta~~.
\ee

\paragraph{Observation} Given an element $\Phi\in {\cal H}$, such that 
$|\Phi|$ is {\em odd}, 
consider the operator:
\be
A_\Phi u:=[\Phi,u]_\bullet~~. 
\ee
It is clear that $A_\Phi$ 
acts as an odd graded derivation 
of the total boundary product $\bullet$. It is also easy to check (upon 
using invariance of the bilinear form) that $A_\Phi$ acts as derivation of 
$\langle .,.\rangle$:
\be
\langle A_\Phi u, v\rangle+(-1)^{|u|}\langle u, A_\Phi v\rangle=0~~.
\ee
Upon combining this with the definition of $h$ and the property $c^2=id$,
one obtains that the Hermitian conjugate 
of $A$ with respect to $h$ has the form:
\be
A_\Phi^\dagger u=(-1)^{|u|}cA_\Phi cu~~.
\ee
In particular, if $\phi\in {\cal H}^1$ is a shift of the background
satisfying the equations of motion $d\phi+\phi\bullet \phi=0$, then 
the shifted differential $d_\phi=d+[\phi,\cdot]_\bullet=d+A_\phi $
is again a degree one derivation of the boundary algebra $({\cal H}, \bullet)$,
whose Hermitian conjugate has the form:
\be
d_\phi^\dagger u =d^\dagger u+A^\dagger_\phi u=
d^\dagger u+(-1)^{|u|}c[\phi,cu]=(-1)^{|u|}cd_\phi cu~~.
\ee
This shows invariance of our formalism with respect to shifts of the 
string vacuum. In particular, all statements of this section apply to 
backgrounds $A$ which need not be diagonal.

\subsubsection{Hodge decomposition, invertibility properties and dualities
on cohomology\label{invertibility}}

As in \cite{superpot}, 
we consider the Hodge decomposition 
${\cal H}=Im d \oplus Im d^\dagger \oplus K$, where:
\bea
\label{hodge}
K=ker d\cap ker d^\dagger=ker \Delta~~&,&~~{\cal H}=
K\oplus im d\oplus im d^\dagger~~,\\
ker d=K\oplus im d~~~~~~~~~~~~~~&,&~~ ker d^\dagger=K\oplus im d^\dagger,
\eea
and the propagator
$U=\frac{1}{d}\pi_d=d^\dagger
\frac{1}{H}(\pi_{d}+\pi_{d^\dagger})=d^\dagger\frac{1}{H}\pi_{d}=
\frac{1}{H}d^\dagger$ (with $H=d^\dagger d+dd^\dagger$) associated 
to the gauge $d^\dagger \phi=0$, where 
$\pi_d$ and $\pi_{d^\dagger}$ are the orthogonal projectors on 
$im d$ and $im d^\dagger$. 
Physical states of our string field theory correspond to elements of 
$K^1$ (the subspace of degree one states lying in $K$):
\be
H^1_d({\cal H})\approx K^1~~.
\ee
This isomorphism depends on the choice of metrics on $L$ and ${\bf E}$.

The restriction of $d$ to the orthogonal complement of its kernel
maps  $(ker d)^\perp=im d^\dagger$ onto $ im d$. This gives a 
bijection $d:im d^\dagger \stackrel{\approx}{\rightarrow} im d$. Since 
$c$ maps $im d$ into $im d^\dagger$ and viceversa, it follows that 
$cd$ and $dc$ give automorphisms of the subspaces $im d^\dagger$ and $im d$, 
respectively. We also note that $c$ preserves the subspace $K$, and 
intertwines its components according to:
\be
c(K^j)=K^{3-j}~~.
\ee
In particular, $c$ induces an isomorphism between $H^j_d({\cal H})$ and 
$H^{3-j}_d({\cal H})$, which we will loosely call `Poincare duality' (even though it 
is an isomorphism between two cohomology groups, rather than between homology and 
cohomology). On the other hand, the non-degenerate bilinear form 
$\langle . , . \rangle$ gives canonical (and metric-independent) isomorphisms:
\be
\label{dual}
H^{3-j}_d({\cal H})\approx H^j_d({\cal H})^*~~.
\ee
The isomorphism induced by $c$ results from this upon composing with the 
identification induced by $h$ between $H^j_d({\cal H})$ and 
its dual.

\section{Example: Topological D-brane pairs of unit relative grade in 
a scalar background\label{example}}

Consider a D-brane pair $(a,b)$ such that $\phi_a:=0$ and $\phi_b=1$. 
In this case, the underlying graded bundle is ${\bf E}=E_a\oplus E_b$, where 
$E_a$ and $E_b$ are the flat bundles underlying the D-branes. 
Throughout this section, we assume that the flat connections $A_a$ 
and $A_b$ are unitary with respect to some choice of 
Hermitian metrics on $E_a$ and $E_b$, which we shall use in order to 
build the metric $h$ on ${\cal H}$. This assumption will be necessary 
for to arrive at a particularly simple form of the deformed Laplacian.

The space ${\cal H}^k$ of degree $k$ elements in ${\cal H}$ is 
the space of sections of the bundle
${\cal V}^k=\Lambda^k(T^*L)\otimes End(E_a)\oplus \Lambda^k(T^*L)\otimes End(E_b)
\oplus \Lambda^{k-1}(T^*L)
\otimes Hom(E_a, E_b)\oplus \Lambda^{k+1}(T^*L)\otimes Hom(E_b, E_a)$. Its elements 
can be arranged as a matrix of bundle-valued forms:
\be
\label{u}
u=\left[\begin{array}{cc}
u_k&u_{k+1}\\
u_{k-1}&{\hat u}_k
\end{array}\right]~~,~~{\rm for}~~|u|=k~~,
\ee
where the subscript denotes form rank and the bundle 
components of $u_k=u_{aa}, {\hat u}_k=u_{bb}, u_{k+1}=u_{ba}$ and 
$u_{k-1}=u_{ab}$ are morphisms 
from $E_a$ to $E_a$, $E_b$ to $E_b$, $E_b$ to $E_a$ and $E_a$ to $E_b$ respectively. 
With this convention, 
the boundary product agrees with matrix multiplication:
\be
(u\bullet v)_{\alpha\beta}=u_{\alpha\gamma}\bullet v_{\gamma\beta}~~
\ee
for all $\alpha, \beta\in \{a,b\}$. 

We are interested in backgrounds of the form:
\be
\label{phi}
\phi=\left[\begin{array}{cc} 0&0\\\phi_0&0\end{array}\right]~~,~~|\phi|=1~~,
\ee
where $\phi_0$ is a
zero-form valued in the bundle $Hom(E_a, E_b)$. In this case, the 
equations of motion $d\phi+\phi\bullet \phi=0$ reduce to $d\phi_0=0$, which 
means that $\phi_0$ is a covariantly-constant section of $Hom(E_a,E_b)$. 
Such backgrounds were also considered in 
\cite{bv}, where we discussed acyclic composite formation 
under some (rather stringent) topological assumptions.
In this section, we generalize that result 
by removing all such restrictions, and study other 
aspects of this system.  

\subsection{The deformed Laplacian\label{deformedlap}} 

Let us find the Laplacian 
$\Delta_\phi=d_\phi d^\dagger_\phi+d^\dagger_\phi d_\phi$ in the 
background (\ref{phi}).  
It turns out that this 
operator has a particularly simple form for our class of backgrounds.
To describe this, we define a {\em graded anticommutator} through:
\be
\{u,v\}_\bullet =u\bullet v +(-1)^{|u||v|}v\bullet u~~{\rm~for~}u,v\in {\cal H}~~.
\ee
This quantity, which differs from the graded commutator 
$[u,v]_\bullet =u\bullet v -(-1)^{|u||v|}v\bullet u$ by the 
middle sign factor, has the graded {\em symmetry} property:
\be
\{u,v\}_\bullet =(-1)^{|u||v|}\{v,u\}_\bullet~~.
\ee
It is shown in Appendix A that the deformed Laplacian can be written as
\footnote{This expression for $\Delta_\phi$ is only valid for
the particular class of backgrounds considered in this section.}:
\be
\label{Delta}
\Delta_\phi u=\Delta u+\{[\phi,\phi^\dagger]_\bullet ,u\}_\bullet ~~,
\ee
where $\phi^\dagger=\left[\begin{array}{cc} 0&\phi_0^\dagger
\\0&0\end{array}\right]$ is the Hermitian conjugate of $\phi$. 
Note that $|\phi^\dagger|=-1$ and:
\bea
[\phi,\phi^\dagger]_\bullet&=&\phi\bullet \phi^\dagger +\phi^\dagger \bullet \phi=
\phi\circ \phi^\dagger +\phi^\dagger \circ \phi=\{\phi, \phi^\dagger\}~~,\\
\{[\phi,\phi^\dagger]_\bullet,u\}_\bullet &=&[\phi,\phi^\dagger]_\bullet
\bullet u+u\bullet [\phi,\phi^\dagger]_\bullet =\{\phi,\phi^\dagger\}\circ u+
u \circ \{\phi,\phi^\dagger\}\nn~~,
\eea
where $\circ$ denotes composition of fiber morphisms and $\{.,.\}$ is the 
usual anticommutator taken with respect to this composition. 
Consider the operator $A_\phi u:=[\phi, u]_\bullet$. Then it is shown in 
Appendix A that $A_\phi^\dagger u=\{\phi^\dagger, u\}_\bullet$.
It follows that the last term of (\ref{Delta}) has the form:
\be
\{[\phi, \phi^\dagger]_\bullet , u\}_\bullet =
(A^\dagger_\phi A_\phi +A_\phi A_\phi^\dagger)~u~~,
\ee
and thus $\Delta_\phi$ is a sum of three non-negative operators:
\be\Delta_\phi=\Delta+A^\dagger_\phi A_\phi +A_\phi A_\phi^\dagger~~.
\ee

Upon noticing that
$[\phi,\phi^\dagger]_\bullet=
\left[\begin{array}{cc} \phi_0^\dagger \circ \phi_0&0\\0&\phi_0\circ 
\phi_0^\dagger\end{array}\right]$, we use (\ref{Delta}) to 
obtain the component form:
\be
\label{Delta_comp}
\Delta_\phi u=
\left[\begin{array}{cc} 
\Delta u_k+u_k\circ \phi_0^\dagger \circ \phi_0+
\phi_0^\dagger \circ \phi_0\circ u_k
& 
\Delta u_{k+1}+u_{k+1}\circ \phi_0\circ \phi_0^\dagger+
\phi_0^\dagger \circ \phi_0\circ u_{k+1}
\\
\Delta u_{k-1}+u_{k-1}\circ \phi_0^\dagger \circ \phi_0+
\phi_0\circ \phi_0^\dagger\circ u_{k-1}
&
\Delta {\hat u}_k+{\hat u}_k\circ \phi_0\circ \phi_0^\dagger+
\phi_0 \circ \phi_0^\dagger\circ {\hat u}_k
\end{array}\right]~~,
\ee
where $u\in {\cal H}^k$ is an element of 
the form (\ref{u}) and $\Delta$ acting on its components stands for the 
Laplacian on bundle-valued forms, coupled to the flat connection in the appropriate 
bundle. 

\subsection{Harmonic states}

Let us look for harmonic elements of ${\cal H}$, i.e. 
solutions to the equation:
\be
\Delta_\phi u=\Delta u+(A_\phi^\dagger A_\phi+A_\phi A^\dagger_\phi)u=0~~.
\ee
Since each of the operators 
$\Delta$, $A_\phi^\dagger A_\phi$ and $A_\phi A^\dagger_\phi$ is
non-negative with respect to the Hermitian scalar product $h$,
this equation is equivalent with:
\be
\label{harmonicity}
\Delta u = 0~~,~~ 
A_\phi u = 0\Leftrightarrow [\phi,u]_\bullet =0~~,~~
A^\dagger_\phi u = 0\Leftrightarrow \{\phi^\dagger , u\}_\bullet =0~~.\nn
\ee
For an element $u$ of worldsheet degree $|k|$ (equation (\ref{u})), 
the last two conditions take the form:
\bea
\label{hs}
&&u_{k-1}\circ \phi^\dagger_0=\phi^\dagger _0\circ u_{k-1}=0~~\nn\\
&&u_{k+1}\circ \phi_0=\phi_0\circ u_{k+1}=0~~\nn\\
&&\phi_0\circ u_k={\hat u}_k\circ \phi_0~~\\
&&\phi^\dagger _0\circ {\hat u}_k=-u_k\circ \phi^\dagger_0~~.\nn
\eea
Let $K(\phi_0)=ker \phi_0$ and $I(\phi_0)=im \phi_0$ 
be the kernel and image of the bundle morphism $\phi_0:E_a\rightarrow E_b$. 
$K(\phi_0)$ and $I(\phi_0)$ are subbundles of $E_a$ and $E_b$ respectively. 
We also consider the orthogonal complement $I^\perp(\phi_0)$ of $I(\phi_0)$
in $E_b$. Note that $ker(\phi_0^\dagger)=I^\perp(\phi_0)$.
The condition $d\phi_0=0$ 
can be used to check that 
that these subbundles are preserved by covariant differentiation. 
This implies:
\bea 
&&d\Omega^*(L, End(K(\phi_0))\subset \Omega^{*+1}(L, End(K(\phi_0)))~~\nn\\
&&d\Omega^*(L, End(I^\perp(\phi_0))\subset 
\Omega^{*+1}(L, End(I^\perp(\phi_0)))~~\nn\\
&&d\Omega^*(L, Hom(K(\phi_0), I^\perp(\phi_0))\subset \Omega^{*+1}(L, 
Hom(K(\phi_0), I^\perp(\phi_0)))~~\\
&&d\Omega^*(L, Hom(I^\perp(\phi_0), K(\phi_0))
\subset \Omega^{*+1}(L, Hom(I^\perp(\phi_0), K(\phi_0)))~~.\nn
\eea

The first two equations in (\ref{hs}) amount to the requirement that 
the fiber components $u_{k+1}$ and $u_{k-1}$ reduce to morphisms between 
$K(\phi_0)$ and $I^\perp(\phi_0)$:
\be
\label{c1}
u_{k-1}\in \Omega^{k-1}(L, Hom(K(\phi_0), I^\perp(\phi_0)))~~,~~
u_{k+1}\in \Omega^{k+1}(L, Hom(I^\perp(\phi_0), K(\phi_0)))~~.
\ee
To solve the last two conditions in (\ref{hs}), we multiply them to the left by 
$\phi_0^\dagger$ and $\phi_0$ respectively and combine the results to obtain:
\be
\phi_0^\dagger \phi_0 u_k+u_k\phi_0^\dagger\phi_0=0~~{\rm and}~~
\phi_0\phi^\dagger_0 {\hat u}_k+{\hat u}_k\phi_0\phi_0^\dagger=0~~.
\ee
Since both $\phi_0^\dagger\phi_0$ and $\phi_0\phi_0^\dagger$ are non-negative 
operators, and since $ker (\phi_0^\dagger \phi_0)=K(\phi_0)$ 
and $ker (\phi_0\phi_0^\dagger)=ker (\phi_0^\dagger)=I^\perp(\phi_0)$, 
this is easily seen to imply: 
\be
\label{c2}
u_k\in \Omega^k(L, End(K(\phi_0)))~~{\rm and}~~
{\hat u}_k\in \Omega^k(L, End(I^\perp(\phi_0)))~~.
\ee

Equations (\ref{c1}) and (\ref{c2}) mean that 
$u$ is a section of the bundle 
$\Lambda^*(T^*L)\otimes End(K(\phi_0)\oplus I^\perp(\phi_0))$.
Finally, the first equation in (\ref{harmonicity}) requires 
that the components $u_k,{\hat u}_k$ and $u_{k-1}$, $u_{k+1}$
be harmonic. We conclude that the space of harmonic elements 
in worldsheet degree $j$ is given by:
\bea
\label{Kk}
K^j_\phi=&&
\Omega^j_{harm}(L,End(K(\phi_0)))\oplus 
\Omega^j_{harm}(L,End(I^\perp(\phi_0)))\oplus~~\\
&&\Omega^{j-1}_{harm}(L,Hom(K(\phi_0), I^\perp(\phi_0)))\oplus
\Omega^{j+1}_{harm}(L,Hom(I^\perp(\phi_0), K(\phi_0)))~~.\nn
\eea 
This situation is described in figure 1.

\begin{figure}[hbtp]
\begin{center}
\scalebox{0.5}{\input{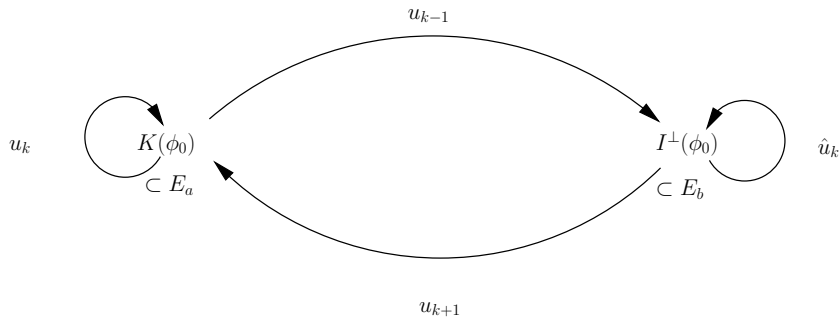}}
\end{center}
\caption{The bundle structure of harmonic states.}
\label{figure1}
\end{figure}

\subsection{General construction 
of acyclic composites\label{acyclic_general}}

A particularly interesting case is $r_a=r_b$, 
and $K(\phi_0)=0$. In this situation, $\phi_0$ is a flat isomorphism from $E_a$ 
to $E_b$, i.e. an isomorphism of $E_a$ and $E_b$ as flat vector bundles. 
Then both $K(\phi_0)$ and $I^\perp(\phi_0)$ coincide with the zero vector 
bundle, and thus the space of zero modes $K^j_\phi$ vanishes for all $j$. 
It follows that such a background is {\em acyclic}, i.e. its 
worldsheet BRST cohomology $H^*_{d_\phi}({\cal H})$ vanishes in all degrees. 
We obtain the following: 

\

{\bf Proposition} 
Suppose $E_a$ and $E_b$ are two flat vector bundles over a 
closed 3-manifold $L$. If $E_a$ and $E_b$ 
are isomorphic as flat bundles, and $\phi_0:E_a\rightarrow E_b$ 
is a flat isomorphism, then the background $\phi_0$ is acyclic. 

\

A similar result was derived in \cite{bv} under (much) more restrictive 
assumptions.
The proposition discussed above removes the limitations of the argument 
of \cite{bv}. This result {\em proves} that 
shifting the grade of a topological D-brane by one unit 
can be viewed as 
transforming the brane into its `topological antibrane', 
irrespective of the topology of the cycle $L$. 
It also proves that 
topological brane-antibrane pairs of the A-model can annihilate at least in 
the large radius limit of a Calabi-Yau compactification.

\subsection{Count of virtual zero modes}

Hodge theory for $d_\phi$ gives isomorphisms 
$H^j_{d_\phi}({\cal H})=K^j$, so that $h_k:=dim_\C H^j({\cal H})=
dim_\C K^j$. On the other hand, Hodge theory on each of the 
subbundles involved in (\ref{Kk}) relates the dimension of the 
associated space of harmonic forms to the corresponding cohomology.
We obtain:
\bea
h_k=&&h^k_{d_{aa}}(L,End(K(\phi_0)))+ 
h^k_{d_{bb}}(L,End(I^\perp(\phi_0)))+~~\\
&&h^{k-1}_{d_{ab}}(L,Hom(K(\phi_0), I^\perp(\phi_0)))+
h^{k+1}_{d_{ba}}(L,Hom(I^\perp(\phi_0), K(\phi_0)))~~,\nn
\eea
where we used the fact that the original flat connection $A$ is a direct 
sum of connections $A_a$ and $A_b$ in the bundles $E_a$ and $E_b$, 
and thus $d$ reduces to the operator $d_{\alpha\beta}$ when restricted to 
$\Omega^*(L, Hom(E_\alpha, E_\beta))$. Here $d_{\alpha\beta}$ is the de Rham
differential twisted by the connection induced 
on $Hom(E_\alpha,E_\beta)$ by $A_\alpha$ and $A_\beta$.

A particularly simple case arises when the flat connections 
$A_\alpha$ are the trivial (i.e. their holonomies are trivial 
around all one-cycles of $L$). In this case, 
one obtains:
\be
h_k=b_k(L)[(rk K(\phi_0))^2+ (rk I^\perp(\phi_0))^2]+
[b_{k-1}(L)+b_{k+1}(L)]rk K(\phi_0)rk I^\perp(\phi_0)~~,
\ee
where $b_j(L)$ is the $j$-th Betti number of $L$ (we define $b_j(L)$ to be 
zero unless $j=0\dots 3$). The rank theorem applied to $\phi_0$ gives
\be
rk I^\perp(\phi_0)=\delta+\Delta r~~,
\ee
where $\delta=rk K(\phi_0)$ is the defect of $\phi_0$ and $\Delta r:=r_b-r_a$. 
Using $b_0(L)=1$ and $b_{3-j}(L)=b_j(L)$, we obtain
$h_j=(b_{j-1}(L)+2b_j(L)+b_{j+1}(L))\delta(\delta+\Delta r)+ 
b_j (L)(\Delta r)^2$, i.e.:
\bea
\label{hks}
&&h_{-1}=h_4=\delta(\delta+\Delta r)~~\\
&&h_0=h_3=(2 +b_1(L))h_{-1}+(\Delta r)^2~~\nn\\
&&h_1=h_2=(1+3b_1(L))h_{-1}+b_1(L)(\Delta r)^2~~.\nn
\eea
Note the identities $h_{3-j}=h_{j}$, which follow from 
`Poincare duality' for $H^*_{d_\phi}({\cal H})$. 
In particular, $h_{-1}$ and the virtual dimension 
$h_1$ are strictly increasing functions
of $\delta$. 
Moreover, it is clear that $h_{-1}=h_4$ determines $\delta$ uniquely:
\be
\delta=\frac{1}{2}[\sqrt{(\Delta r)^2+4 h_{-1}} -\Delta r]~~.
\ee
The defect $\delta$ 
stratifies the moduli space in the directions accessible 
by turning on $\phi_0$. 
For the original background, one has $\phi_0=0$, so $\delta=r_a$
and the virtual dimension $h_1=(1+3b_1(L))r_a r_b +b_1(L)(\Delta r)^2=
r_a r_b +b_1(L)(r_a^2+r_b^2+r_a r_b)$ 
are at their maximum values. 
As we vary $\phi_0$, we pass through 
strata of lower virtual dimension, according to the decreasing rank 
of its kernel. 
Since $rk I^\perp(\phi_0)=\delta+\Delta r\geq 0$, the minimal value of 
$\delta$ is 
$max(0, -\Delta r)$, for which $h_{-1}$ vanishes and $h_1$ attains its 
minimum, equal to $b_1(L)(\Delta r)^2$. This gives a total of 
$1+min(r_a, r_b)$ strata, whose virtual dimensions form the 
histogram in figure 2.

\begin{figure}[hbtp]
\begin{center}
\scalebox{0.5}{\input{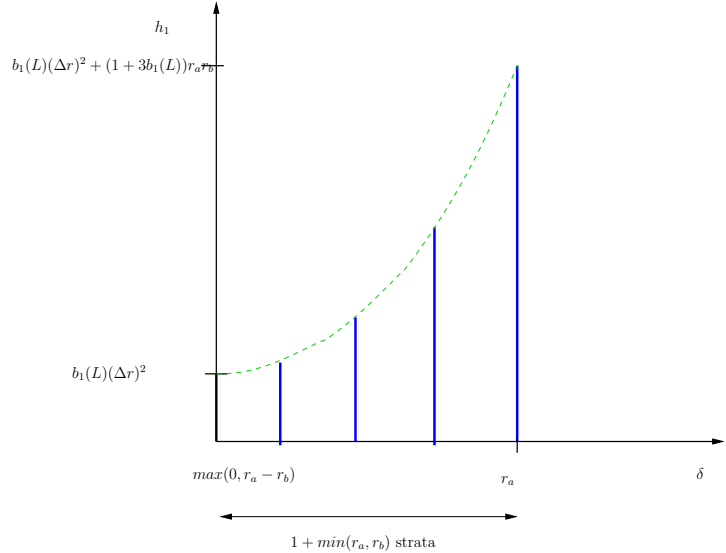}}
\end{center}
\caption{Virtual dimensions of the various strata.}
\label{figure2}
\end{figure}

As an example, consider the case $r_a=r_b:=r$, so $\Delta r=0$ and 
$\delta\in \{0,...,r\}$. In this case, $h_1=(1+3b_1(L))\delta^2$. 
For $L$ a rational homology sphere, one has $b_1(L)=0$ and 
$h_1=\delta^2$. Thus the trivial $\phi_0=0$ lies in 
a component of virtual dimension equal to $r^2$, while an invertible 
$\phi_0$ is an isolated background (belongs to a component 
of vanishing virtual dimension); the latter describes the acyclic composite. 
For a torus $L=T^3$, 
one has $b_1(L)=3$ and $h_1=10\delta^2$. For $r=1$ (a pair of singly wrapped 
graded branes), one obtains two strata $\delta=0,1$, of virtual dimensions $0$ 
and $10$. The first stratum contains a single point, namely the acyclic
composite. The second stratum results by condensation of spacetime fields 
associated with boundary condition changing states.
As we shall see in Section 5, the correct picture 
of this stratum is quite different, due to the obstructed character of 
some deformations. 
A similar analysis, though much more complicated, can be given 
for deformations along $\phi_2$.

\section{The partition function and the holomorphic potential}

\subsection{Introduction and physical interpretation} 

In this section, we give 
an alternate description of the moduli space 
${\cal M}$ in terms of a potential 
for the `low energy modes' which is invariant under certain symmetries. 
Since the discussion is somewhat technical, we start 
with a short explanation of the origin of $W$ and 
its physical meaning. Many ideas can 
be traced back to the work of \cite{Witten_CS}. 

To study the dynamics of moduli, one 
fixes a classical vacuum, to be taken as a starting point for 
the perturbation expansion. Such a vacuum is a degree one 
flat graded connection 
$A$ in the graded bundle ${\bf E}$, i.e. a point in the 
classical moduli space ${\cal M}$. Fixing a vacuum leads to spontaneous
symmetry breaking of the gauge group ${\cal G}$ down to the subgroup 
$G_A$ which stabilizes $A$. The quotient ${\cal G}/G_A$ acts 
on $A$, producing its gauge orbit ${\cal O}_A$. Fluctuations tangent to this 
orbit can be viewed as the Goldstone bosons of broken gauge invariance, 
while those `orthogonal' to it can be divided into massless modes 
tangent to ${\cal S}$ (which give moduli) and massive modes orthogonal 
to ${\cal S}$\footnote{${\cal S}$ was defined in Section 2.6.}. 

Consider the infinitesimal situation. A fluctuation $u\in {\cal H}^1$
around $A$ satisfies the classical equations of 
motion if $d_Au=0$, which means that $u$ belongs to the tangent space 
$T_A{\cal S}$. This space can be identified with 
$(ker d_A)^1=ker (d_A:{\cal H}^1\rightarrow {\cal H}^2)$. 
Infinitesimal gauge transformations $A\rightarrow A-d_A\alpha$ 
(with $\alpha\in {\cal H}^0$) stabilize $A$ if and only if:
\be
d_A \alpha=0~~.
\ee
The infinitesimal action is trivial if $\alpha$ belongs to the image of 
$d_A$. Hence the Lie algebra of $G_A$  
can be identified with the cohomology group $H^0_{d_A}({\cal H})$ 
(endowed with the induced Lie bracket, which is well-defined by virtue of 
the derivation property of $d_A$). In particular, $G_A$ is a 
{\em finite-dimensional} Lie group. As a vector space, the Lie algebra of 
${\cal G}/G_A$ can be identified with ${\cal H}^0/(ker d_A)^0$. 
On the other hand, the space  $T_A{\cal O}_A$ of Goldstone modes 
coincides with  
$(im d_A)^1:=im (d_A:{\cal H}^0\rightarrow {\cal H}^1)$. 
The space of moduli is given by 
$(ker d_A)^1/(im d_A)^1=H^1_{d_A}({\cal H})$, while  
`massive modes' are described by ${\cal H}^1/(ker d_A)^1$ (figure 3).

\begin{figure}[hbtp]
\begin{center}
\scalebox{0.4}{\input{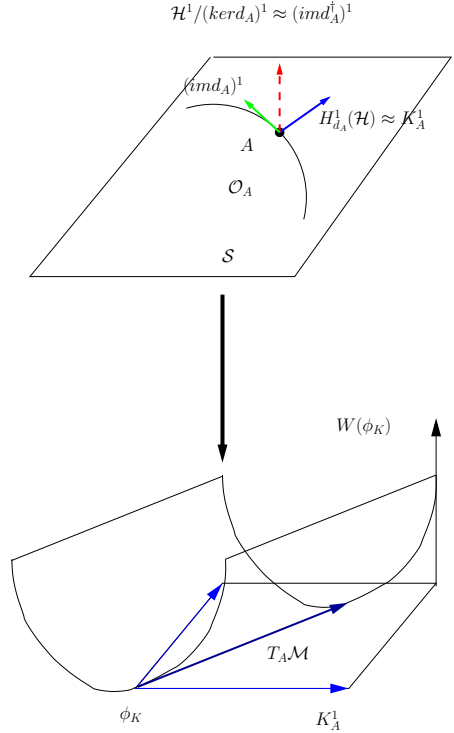}}
\end{center}
\caption{Massive, harmonic and Goldstone modes. After gauge fixing and 
integrating out the massive modes, one obtains a potential $W$ for the 
harmonic modes, whose critical set characterizes the 
true moduli.}
\label{figure3}
\end{figure}

To eliminate the Goldstone modes, one must pick a gauge-fixing condition. 
We follow \cite{superpot} by choosing the Lorenz gauge:
\be
\label{gc}
d_A^\dagger u=0~~,
\ee
with $d_A^\dagger$ defined as in Section 2. 
Using the Hodge decomposition ${\cal H}=im d_A^\dagger\oplus im d_A\oplus K_A$,
we can  
identify the Goldstone modes with $(im d_A)^1$, the moduli with $K^1_A$ 
and the `massive modes' with $[(ker d_A)^1]^\perp=(im d_A^\dagger)^1$. 
We also identify the underlying vector spaces of the 
Lie algebras of $G_A$ and ${\cal G}/G_A$ with
$K^0$ and $(im d_A)^0$. These identifications
need not respect the product structure 
on ${\cal H}$, and therefore
need not respect\footnote{This follows from 
the basic observation that the product of two Harmonic forms may fail to be be 
harmonic.} the Lie structure on ${\cal H}^0$
(in  particular, the 
commutator of two elements of $K^0$ need not belong to $K^0$).

Moreover, the operator 
$d^\dagger_A$ need  {\em not} obey the usual property 
$d^\dagger_A  Ad_g(u)=Ad_g(d^\dagger_A u)$ (for $g\in G_A$). 
Thus, the gauge condition (\ref{gc}) may not be invariant with respect to the 
adjoint action of $G_A$. As a result, this action 
generally mixes massive modes and moduli.

An effective description of moduli is obtained 
by integrating out all massive modes, 
which will produce a potential defined on the space $K^1$ of 
linearized deformations. The potential is defined formally as follows. 
Since the gauge condition (\ref{gc}) eliminates the Goldstone modes $im d$, it 
allows us to restrict to the subspace $(ker d^\dagger)^1=(im d^\dagger)^1\oplus K^1$.
Upon decomposing $\phi\in (ker d^\dagger)^1$ as:
\be
\phi=\phi_K\oplus \phi_M~~{\rm with}~~\phi_K\in K^1~{\rm~and~}~~\phi_M\in 
(im d^\dagger)^1~~,
\ee
we define an all-order potential for the massive modes through:
\be
e^{-\frac{i}{\hbar}W_{full}(\phi_K)}=
\int{{\cal D}[\phi_M]e^{-\frac{i}{\hbar}S(\phi_K\oplus \phi_M)}}~~.
\ee
This equation defines the potential to all loop orders. 
The potential is nontrivial due to the existence of cubic interactions 
between harmonic and `massive' modes\footnote{To reach this equation, 
we noticed 
that $\langle \phi_K , d\phi_M\rangle =\langle d\phi_K, \phi_M\rangle=0$,  
since $d\phi_K=0$.}:
\bea
\label{split}
S(\phi_K\oplus \phi_M)&=&\frac{1}{2}\langle \phi_M, d\phi_M \rangle+
\frac{1}{3}\langle \phi_M, \phi_M\bullet \phi_M\rangle   \\
&+&
\langle \phi_K, \phi_M\bullet \phi_M\rangle+
\langle \phi_M, \phi_K\bullet \phi_K\rangle+
{1\ov 3}\langle\phi_K,\phi_K\bullet\phi_K\rangle~~. \nn
\eea
Thus\footnote{The path is normalized by
$\int{{\cal D}[\phi_M]e^{-\frac{i}{\hbar}S(\phi_M)}}=1$.}:
\be
e^{-\frac{i}{\hbar}W_{full}(\phi_K)}=
e^{-{i\ov 3\hbar}\langle\phi_K,\phi_K\bullet\phi_K\rangle}
\int{{\cal D}[\phi_M]
e^{-\frac{i}{\hbar}S(\phi_M)}e^{-\frac{i}{\hbar}
\left(\langle \phi_K, \phi_M\bullet \phi_M\rangle+
\langle \phi_M, \phi_K\bullet \phi_K\rangle\right)}}~~,
\label{fullpot}
\ee
which gives a perturbative series for $W_{full}$ upon expanding the 
last exponential.
This leads to Feynman graphs built out of the vertices and massive propagator 
depicted in figure 4. 
If $W$ denotes the tree-level approximation to $W_{full}$, 
then its expansion has the form discussed in \cite{superpot}. 

\begin{figure}[hbtp]
\begin{center}
\scalebox{0.5}{\input{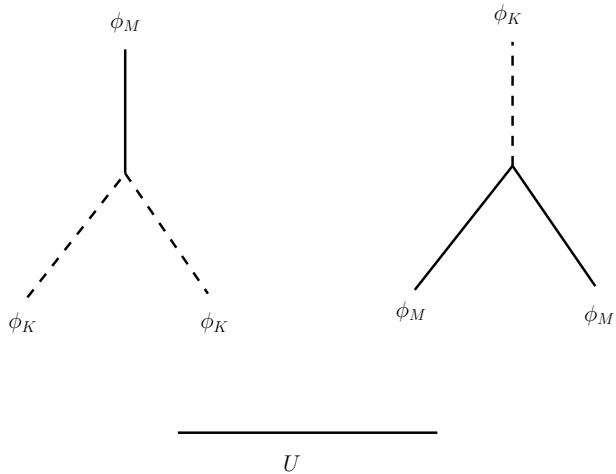}}
\end{center}
\caption{Physical vertices and propagator for the perturbative expansion of $W$. 
Dashed lines represent massless  modes. Beyond tree level, the perturbation 
expansion also involves vertices and propagators for 
ghost and antighosts, which are not shown in the figure.}
\end{figure}

Since the adjoint action of $G_A$ mixes moduli and massive modes, 
integrating out the later induces certain symmetries of $W$. 
The precise form of these symmetries can be determined 
directly from the potential. As in \cite{superpot}, this 
will give a local description of the 
moduli space as the quotient of the critical set of $W$ through 
these symmetries.

\subsection{On justifying the Lorenz gauge and 
the decoupling of ghosts in the tree-level potential}

Our discussion of the potential was somewhat naive, since we did not 
attempt to give a complete treatment of the gauge-fixing procedure; 
in particular, we did not discuss the ghost/antighost contributions to the gauge-fixed
action, and their role in the perturbative expansion of $W$
\footnote{The case of ungraded D-branes involves similar issues; 
in that situation, the direct approach of \cite{superpot} is justified 
due to results of \cite{AS1}.}.
 
In fact, since the gauge algebra of the theory (\ref{action}) is generally reducible, 
it is far from clear that the approach outlined above is indeed correct. 
For a rigorous treatment, 
one must {\em show} that (\ref{gc}) is a well-defined gauge-fixing 
condition, and understand the role of ghosts and antighosts in the perturbative 
expansion.
A complete analysis of this issue requires the  
Batalin-Vilkovisky formalism and is performed in \cite{gf}. There it is 
showed that:

\

(1)The gauge-fixing condition defined by (\ref{gc}) is consistent, 
and results from an appropriate gauge-fixing fermion.  In particular, 
the relevant propagators can be determined by the method of \cite{superpot}. 

\

(2)The {\em tree-level} potential $W$ receives no ghost/antighost contributions, 
and can be computed from Feynman 
diagrams involving only the ingredients shown in figure 4. 

\

These conclusions result from a somewhat technical BV analysis of gauge-fixing, 
starting with the master action of our systems, which was constructed in 
\cite{sc} and further discussed in \cite{bv}.

\subsection{Expansion of the tree-level potential}

The results of \cite{gf} assure us that we can use 
the gauge condition (\ref{gc})
and neglect the issue of ghost contributions, {\em as long as} 
we are interested in tree-level diagrams only. Since the algebraic framework of 
\cite{superpot} is also satisfied (as showed in Section 2), we can 
apply the construction of that paper and carry over its results.

As explained in \cite{superpot}, 
tree-level amplitudes of massless states $u_1..u_n\in K^1$ can be written in the 
form:
\be
\label{amplitudes}
\langle \langle u_1\dots u_n\rangle \rangle^{(n)}_{tree}=
\langle u_1, r_{n-1}(u_2\dots u_n)\rangle~~,
\ee
where the scattering products $r_n:K^{\otimes n}\rightarrow K$ 
$(n\geq 2)$ are recursively defined as follows: 

\

1. We first define multilinear maps 
$\lambda_n:{\cal H}^n\rightarrow {\cal H}$ 
through $\lambda_2=\bullet$ and the recursion relation:
{\footnotesize \bea
\label{rec}
\lambda_n(u_1,\dots ,u_n)&=&(-1)^{n-1}
(U\lambda_{n-1}(u_1,\dots ,u_{n-1}))\bullet u_n-(-1)^{n|u_1|}
u_1\bullet (U\lambda_{n-1}(u_2,\dots ,u_n))-\nn\\
&&\sum_{\tiny \begin{array}{c}k+l=n\\ k,l\ge 2\end{array}}
(-1)^{k+(l-1)(|u_1|+\dots +|u_k|)}
(U\lambda_k(u_1,\dots ,u_k))\bullet (U\lambda_l(u_{k+1},\dots ,u_n))~~,
\eea}
for $u_1\dots u_n$ in ${\cal H}$. 

\

2. The products $r_n$ are then given by:
\be
\label{rs}
r_n(u_1,\dots ,u_n)=P\lambda_n(u_1,\dots ,u_n)~~,
\ee
for $u_1, \dots  , u_n\in K$. Here $P$ is the orthogonal projector on 
$K$. The propagator $U$ was defined in Subsection 2.7.3.

This description follows from the tree-level Feynman diagrams associated 
with the expansion of $W$. Appendix B gives an alternate 
(but equivalent) justification, 
which follows from the JWKB approximation.
As in \cite{superpot}, the tree-level potential can be expressed as:
\be
\label{superpot}
W[\phi_K]=-\sum_{n\geq 3}{\frac{1}{n}(-1)^{n(n-1)/2}
\langle \langle \phi,\dots ,\phi\rangle \rangle^{(n)}_{tree}}~~,
\ee
where the massless mode $\phi_K$ belongs to $K^1=K\cap {\cal H}^1$.

As discussed in \cite{superpot}, the products (\ref{rs}) define an algebraic 
structure on ${\cal H}$ known as an $A_\infty$ -{\em algebra}. Since there
is no first order product $r_1$, this $A_\infty$ algebra is {\em minimal}. 
Moreover, it is {\em quasi-isomorphic} with the differential graded
algebra $({\cal H}, d, \bullet)$, if the later is viewed as an $A_\infty$ 
algebra whose higher products vanish. In particular, changing the metrics
on $L$ and $E_n$ leads to $A_\infty$ products which differ by 
quasi-isomorphisms; this amounts to a change of variables in the potential $W$. 
It can also be shown that 
$r_n$ satisfy the cyclicity constraints:
\be
\label{cc2}
\langle u_1, r_n(u_2\dots u_{n+1})\rangle=(-1)^{n(|u_2|+1)}
\langle u_2, r_n(u_3\dots u_{n+1},u_1)\rangle~~.
\ee

\subsection{An alternate description of the moduli space}

As in \cite{superpot}, 
the algebraic structure obeyed by $r_n$ implies that 
the moduli space ${\cal M}$ of (\ref{action})
is locally isomorphic\footnote{More precisely, the associated deformation 
functors are equivalent.} with a moduli space ${\cal M}_W$ constructed
from the potential as follows. 
Consider the critical point condition:
\be
\label{mc_W}
\frac{\partial W}{\partial \phi}(\phi)=0
\Leftrightarrow \sum_{n\geq 2}{(-1)^{n(n+1)/2}r_n(\phi^{\otimes n})}=0~~,
\ee
which can also be written in the form:
\be
\label{mc_L}
\sum_{n\geq 2}{\frac{(-1)^{n(n+1)/2}}{n!}m_n(\phi^{\otimes n})}=0~~,
\ee
upon defining the new products:
\be
\label{commutator}
m_n(u_1\dots u_n)=\sum_{\sigma\in S_n}{\chi(\sigma, u_1\dots u_n)
r_n(u_{\sigma(1)}\dots 
u_{\sigma(n)})}~~,
\ee
where $\chi(\sigma, u_1\dots u_n)$ is the modified 
Koszul sign (see \cite{superpot}). 
Equation (\ref{mc_W}) and the potential are 
invariant with respect to infinitesimal symmetries of the form:
\be
\label{gauge_W}
\phi\rightarrow \phi'=\phi+\delta_\alpha\phi~, {\rm~with~}~
\delta_\alpha\phi=-\sum_{n\geq 2}{\frac{(-1)^{n(n-1)/2}}{(n-1)!}
m_n(\alpha\otimes \phi^{\otimes n-1})}~~,\nn
\ee
where $\alpha$ is a degree zero element of $K$. Correspondingly, we define 
${\cal M}_W$ to be the moduli space of solutions to (\ref{mc_W}), modulo 
the identifications induced by transformations (\ref{gauge_W}). 
It can be shown that the products (\ref{commutator}) form an $L_\infty$ 
algebra, the so-called {\em commutator algebra} of the $A_\infty$ algebra
$(r_n)$.  In the formulation (\ref{mc_L}), 
(\ref{gauge_W}),
the moduli problem is sometimes known as a `homotopy Maurer-Cartan 
problem' and was studied for example in \cite{Kontsevich_Felder} and 
\cite{Merkulov_infty}. We refer the reader to \cite{superpot} for 
an overview of the relevant results.

\paragraph{Observations} 

(1) In the case of ungraded A/B branes, 
the symmetries (\ref{gauge_W}) close to a Lie algebra \cite{superpot}. 
This follows from the property $d^\dagger(\alpha\bullet u)=\alpha\bullet 
d^\dagger u {\rm~for~}\alpha\in K^0 {\rm~and~}u\in {\cal H}$, which holds in 
those models. In the graded case considered in the present paper, 
this property need not hold, 
and the generators of (\ref{gauge_W}) may fail 
to form a Lie algebra. Nonetheless, the 
definition of ${\cal M}_W$ can be formulated geometrically 
in the language of \cite{Kontsevich_Felder}. Since
one may not be able to associate (\ref{gauge_W}) with a Lie group action, 
one cannot always 
interpret $W$ as a superpotential of a standard supersymmetric 
gauge theory which would 
describe the slow  dynamics of graded D-branes. This is why we 
prefer the term `holomorphic potential'. 

(2)The potential $W$ can be viewed as a holomorphic function defined on 
$H^1_{d_A}({\cal H})$. This follows upon choosing a 
basis $e_j$ of the finite-dimensional vector space $K^1$, and writing 
$\phi=\sum_{j=1}^{b}{t_je_j}$, where $t_j$ are the associated complex 
coordinates and $b$ is the complex dimension of 
$K^1\approx H^1_d(End({\bf E}))$. Then $W$ can be written in the form:
\be
W(t)=\sum_{s_1..s_b\geq 0}{w_{s_1...s_b}t_1^{s_1}...t_b^{s_b}}~~,
\ee
where $w_{s_1..s_b}$ is given by a sum of expressions involving the 
value of the product $r_{s_1+..+s_b}$ on the appropriate 
collections of basis vectors. Hence $W$ becomes a (formal) power series 
in the complex coordinates $t_j$, and should induce 
a holomorphic function upon performing the required analysis
of convergence. $t_j$ can be viewed as 
local coordinates on the moduli space ${\cal M}$. 
$W$ is holomorphic since we do not impose an (anti) 
self-adjointness condition on the string field $\phi$.

\section{Application to D-brane pairs on $T^3$}

Let us consider the case of {\em trivial} flat bundles $E_a$ and $E_b$ on 
a 3-torus $L$, in a scalar background $\phi_0$ (see eq. (\ref{phi})). 
Upon trivializing $E_a$ and $E_b$, we can view the components 
of (\ref{u}) as matrix-valued forms. Then the condition $d\phi_0=0$ means
that $\phi_0$ is a constant matrix. 
In this subsection, we discuss the holomorphic potential and effective symmetry 
group in this situation. 

\subsection{Preparations}

For the purpose of gauge-fixing we must pick a metric on $L$, and we  
choose this to be the flat metric which makes it into an orthogonal
torus with coordinates $x^j\in [0,2\pi)$:
\be
ds^2=(dx^1)^2+(dx^2)^2+(dx^3)^2~~.
\ee

Harmonic forms 
on $L$ have constant coefficients in these coordinates:
\bea
\label{Omega_harm}
\Omega^0_{harm}(L)&=&\{\omega_0|\omega_0=ct\}~~,\nn\\
\Omega^1_{harm}(L)&=&\{\omega_1 dx^1+\omega_2dx^2 +\omega_3 dx^3|\omega_j=ct\}
~~,\nn\\
\Omega^2_{harm}(L)&=&\{\omega_{12}dx^1 \wedge dx^2+ 
\omega_{23}dx^2 \wedge dx^3+\omega_{31}dx^3 \wedge dx^1|\omega_{ij}=ct\}~~,
\nn\\
\Omega^3_{harm}(L)&=&\{\omega_{123}dx^1\wedge dx^2\wedge dx^3|
\omega_{123}=ct\}~~.
\eea
The Hodge operator acts through:
\be
*(dx^{i_1}\wedge \dots \wedge dx^{i_k})=\frac{1}{(3-k)!}
\epsilon_{i_1\dots i_3}dx^{i_{k+1}}\wedge\dots\wedge dx^{i_3}~~,
\ee
so that: 
\bea
&&*1=dx^1\wedge dx^2\wedge dx^3~~,~~*(dx^1\wedge dx^2\wedge dx^3)=1~~,\nn\\
&&*dx^1=dx^2\wedge dx^3~~,~~*dx^2=dx^3\wedge dx^1~~,
~~*dx^3=dx^1\wedge dx^2~~,\\
&&*(dx^1\wedge dx^2)=dx^3~~,~~
*(dx^2\wedge dx^3)=dx^1~~,~~*(dx^3\wedge dx^1)=dx^2~~.
\nn
\eea
Recall that $*^2=id$ in three dimensions. 

\subsection{General analysis}

As discussed above, harmonic elements of worldsheet degree $k$ 
have the form (\ref{u}), with the various components constrained to belong 
to the spaces listed in equation (\ref{Kk}). Since in our case 
$A_a$ and $A_b$ are 
trivial connections, these subspaces decompose as:
\bea
\label{Omega_decomp}
&&\Omega^k_{d-harm}(L,End(K(\phi_0)))=\Omega^k_{harm}(L)\otimes 
\Gamma(End(K(\phi_0)))~~\nn\\
&&\Omega^k_{d-harm}(L,End(I^\perp(\phi_0)))=\Omega^k_{harm}(L)\otimes
\Gamma(End(I^\perp(\phi_0)))~~\\
&&\Omega^{k-1}_{d-harm}(L,Hom(K(\phi_0), I^\perp(\phi_0)))
=\Omega^{k-1}_{harm}\otimes \Gamma(Hom(K(\phi_0), I^\perp(\phi_0)))~~\nn\\
&&\Omega^{k+1}_{d-harm}(L,Hom(I^\perp(\phi_0), K(\phi_0)))=
\Omega^{k+1}_{harm}\otimes \Gamma(Hom(I^\perp(\phi_0), K(\phi_0)))~~.\nn
\eea
When combined with knowledge of  
$\Omega^k_{harm}(L)$, this leads to a dramatic simplification. 
Indeed, it follows from (\ref{Omega_harm}) that the total space 
$\Omega^*_{harm}(L)$ of harmonic forms on $L$ is closed with respect 
to the wedge product (i.e. the product of two harmonic forms on the three torus
is also harmonic). On the other hand, it is clear that the 
subbundle $End(K(\phi_0)\oplus I^\perp(\phi_0))$ is closed with respect to 
fiberwise composition.
It follows that the space $K^m=ker \Delta_\phi$ 
of $d_\phi$-harmonic elements 
of worldsheet degree $m$ is closed with respect to the total boundary 
product $\bullet$. In particular, given two elements $u, v\in K$, we 
have $d_{\phi}^\dagger(u\bullet v)=0$, and thus $U(u\bullet v)=0$, where 
$U=\frac{1}{\Delta_\phi}d^\dagger$ is the propagator of the `massive' modes.
This implies that the products $r_n$ (with $u_1..u_n \in K$) of Subsection 4.3
vanish for all $n\geq 3$. Therefore, the potential receives contributions 
only from its cubic term: 
\be
\label{pot_scalar}
W(u)=\frac{1}{3}\langle u, r_2(u,u)\rangle=
\frac{1}{3}\langle u, u\bullet u \rangle=
\frac{1}{3}\int_L{str_{End(K(\phi_0)\oplus I^\perp(\phi_0))}
(u\bullet u \bullet u)}~~,
\ee
where $grade(K(\phi_0))=grade E_a=0$ and 
$grade(I^\perp(\phi_0))=grade E_b=1$ in the supertrace. 

Moreover, the infinitesimal effective symmetries 
$\delta_\alpha u$ ($\alpha\in K^0, u\in K^1$)
reduce to:
\be
\delta_\alpha u=m_2(\alpha\otimes u)=[\alpha, u]_\bullet =ad_\alpha(u)~~,
\ee 
which integrate to the adjoint action of a group $G$
on the subspace $K^1$:
\be
\label{G_scalar}
u\rightarrow e^{ad_\alpha}u~~,~~{\rm for~~}\alpha\in K^0, 
u\in K^1~~. 
\ee
$G$ is the Lie group whose Lie algebra is $(K^0, [.,.]_\bullet)$.
It can be described as the group of elements $g\in K^0$ which are invertible 
with respect to the boundary product $\bullet$:
\be
G=\{g\in K^0|{\rm there~exists~}g^{-1}\in K^0{\rm ~such~that}~g\bullet g^{-1}=
g^{-1}\bullet g=1\}~~.
\ee 
The adjoint action (\ref{G_scalar}) is then given by:
\be
\label{G_adj}
u\rightarrow g\bullet u \bullet g^{-1}~~.
\ee
This form of the $G$-action results from the very simple 
form of Hodge theory on the torus, and of our choice of trivial background
connections\footnote{A similar simplification appears for the ungraded case, 
though for different reasons\cite{superpot}.}.  

The critical set of (\ref{pot_scalar}) is described by the equations:
\be
\label{crit}
\frac{\delta W}{\delta u}=0\Leftrightarrow u\bullet u=0~~, 
\ee
which define a subset ${\cal Z}$ of the space $K^1$. 
The moduli space is locally described by the quotient ${\cal Z}/G$. 

To compute $W$ explicitly, we start with the form 
$u=\left[\begin{array}{cc}u_1&u_2\\u_0&{\hat u}_1
\end{array}\right]$ of degree one harmonic elements.
Substituting in (\ref{pot_scalar}) gives:
\be
\label{W_scalar}
W(u)=\int_{L}{tr_{End(K(\phi_0))}\left[u_1^3+3 u_1u_2u_0\right]}-
\int_{L}{tr_{End(I^\perp(\phi_0))}\left[
{\hat u}_1^3+3 u_2{\hat u}_1 u_0\right]}~~,
\ee
where juxtaposition stands for the wedge product and we used the cyclicity 
property of the trace.

Using the decompositions (\ref{Omega_decomp}), we write:
\bea
\label{utorus}
u_1&=&dx^1 X_1+dx^2 X_2+dx^3 X_3~~\nn\\
{\hat u}_1&=&dx^1 Y_1 +dx^2 Y_2 +dx^3 Y_3~~\\
u_2&=&dx^1\wedge dx^2 Z_{12} +dx^2\wedge dx^3 Z_{23}+
dx^3\wedge dx^1 Z_{31}~~\nn\\
u_0&=&W~~,\nn
\eea
where $X_i, Y_i, Z_{ij}$ and $W$ are {\em constant} sections of the  bundles
$End(K(\phi_0))$, $End(I^\perp(\phi_0))$, $Hom(I^\perp(\phi_0), K(\phi_0))$ 
and $Hom(K(\phi_0), I^\perp(\phi_0))$ respectively 
(since $\phi_0$ is 
constant on $L$, all of these bundles are trivial, and thus $X, Y, Z, W$ 
can be viewed as constant linear operators).
With these notations, the potential (\ref{W_scalar}) takes the form:
\bea
\label{Wtorus}
&&W(u)=(2\pi)^3tr_{End(K(\phi_0))}
\left(X_1[X_2,X_3]+W(X_1Z_{23}+X_2Z_{31}+X_3Z_{12})\right)
\nn\\
&&-(2\pi)^3tr_{End(I^\perp(\phi_0))}
\left(Y_1[Y_2,Y_3]+W(Z_{23}Y_1+Z_{31}Y_2+Z_{12}Y_3)\right)~~.
\eea
while the critical point condition (\ref{crit}) reduces to:
\bea
\label{Z}
&&Z_{12}W+[X_1,X_2]=Z_{23}W+[X_2,X_3]=Z_{31}W+[X_3,X_1]=0~~\nn\\
&&Z_{12}W+[Y_1,Y_2]=Z_{23}W+[Y_2,Y_3]=Z_{31}W+[Y_3,Y_1]=0~~\nn\\
&&WX_1-Y_1W=WX_2-Y_2W=WX_3-Y_3W=0~~\\
&&(X_1Z_{23}-Z_{23}Y_1)+(X_2Z_{31}-Z_{31}Y_2)+(X_3Z_{12}-Z_{12}Y_3)=0~~.\nn
\eea
These equations define an algebraic variety ${\cal Z}$, 
which must be further divided
by the action (\ref{G_adj}) to find the moduli space.

\subsection{The case of singly-wrapped branes on $T^3$}

Let us consider the case $r_a=r_b=1$, with the trivial background $\phi=0$. 
Then $E_a$ and $E_b$ are both given by 
the trivial flat line bundle ${\cal O}_L$, and the reference superconnection
is the trivial flat connection on 
${\bf E}={\cal O}_L^{\oplus 2}$. In this case, the 
component of ${\cal H}$ of worldsheet degree $k$ consists of 
elements:
\be
\label{vk}
v=\left[\begin{array}{cc}v_k&v_{k+1}\\v_{k-1}&{\hat v}_k\end{array}\right]~~,
\ee
where the entries $v_j$ are complex-valued forms on $L$. 
$d_A$ is simply a direct sum of de Rham differentials.

\subsubsection{The low energy symmetry group}

Recall that $G$ is the group of 
invertible elements of the associative algebra $(K^0,\bullet)$. 
Such elements have the form:
\be
g=\left[\begin{array}{cc}g_0 & g_1\\0& {\hat g}_0\end{array}\right]~~,
\ee
where $g_0$ and ${\hat g}_0$ are non-vanishing complex constants and 
$g_1$ is a one-form with constant complex coefficients. The group multiplication
and inverse are given by:
\be
\label{inverse}
g\bullet g'=\left[\begin{array}{cc}g_0g_0' & ~~g_0 g'_1+{\hat g}_0'g_1
\\0& {\hat g}_0{\hat g}_0'\end{array}\right]~~,~~
g^{-1}=\left[\begin{array}{cc}g_0^{-1} & -\frac{1}{g_0{\hat g}_0}g_1
\\0& {\hat g}_0^{-1}\end{array}\right]~~.
\ee
where juxtaposition denotes usual multiplication. 

Those elements $g\in G$ which are close to the identity 
can be parameterized exponentially:
\be
g=e^\alpha_\bullet :=\sum_{n\geq 0}{\frac{1}{n!}\alpha^{\bullet n}}~~,~~\alpha\in K^0~~,
\ee
where $\alpha=\left[\begin{array}{cc} \alpha_0& 
\alpha_1\\0&{\hat \alpha}_0\end{array}
\right]$, $\alpha^{\bullet 0}:=id$ 
and $\alpha^{\bullet n}$ stands for the $n$-fold 
$\bullet$-product of $\alpha$ with itself. 
The series converges because 
$K^0$ is finite-dimensional. 

To compute $e^\alpha_\bullet$, we first note that:
\be
\label{alpha_n}
\alpha^{\bullet n}=\left[\begin{array}{cc}
\alpha_0^n& (\sum_{\tiny \begin{array}{c}i+j=n-1\\i,j\geq 0\end{array}}
{\alpha_0^i{\hat \alpha}_0^j})\alpha_1
\\0& {\hat \alpha}_0^n \end{array}\right]~~,
\ee
where juxtaposition stands for usual multiplication. Equation 
(\ref{alpha_n}) follows by a simple induction argument. Upon using this 
result, we compute:
\be
e^\alpha_\bullet=\left[\begin{array}{cc}
e^{\alpha_0} & S\alpha_1
\\0& e^{{\hat \alpha}_0} \end{array}\right]~~,
\ee
where: 
\be
S:=\sum_{n\geq 1}{\frac{1}{n!}\sum_{
\tiny \begin{array}{c}i+j=n-1\\i,j\geq 0\end{array}}{\alpha_0^i
{\hat \alpha}_0^j}}
~~.
\ee
To simplify this, we define $t:=\frac{\alpha_0}{\hat \alpha_0}$ and 
(assuming $t\neq 1$)
use the identity $\sum_{i=0}^{n-1}{t^i}=\frac{~1-t^n}{1-t}$ to obtain:
\be
S=\sum_{n\geq 1}{\frac{{\hat \alpha}_0^n}{n!}\frac{1-t^n}{1-t}}=
\frac{1}{{\hat \alpha}_0-\alpha_0}
\sum_{n\geq 1}{\frac{{\hat \alpha}_0^n-\alpha_0^n}{n!}}=
\frac{e^{\alpha_0}-e^{{\hat \alpha}_0}}{\alpha_0-{\hat \alpha}_0}~~.
\ee
The final equality also holds
for $\alpha\neq \alpha_0$, if the right hand side is interpreted as 
a limit:
\be
S|_{\alpha_0={\hat \alpha}_0}=\lim_{{\hat \alpha}_0\rightarrow \alpha_0}
{\frac{e^{\alpha_0}-e^{{\hat \alpha}_0}}{\alpha_0-{\hat \alpha}_0}}=
e^{\alpha_0}~~.
\ee
We conclude that:
\be
e^\alpha_\bullet =\left[\begin{array}{cc}
e^{\alpha_0} & 
\frac{e^{\alpha_0}-e^{{\hat \alpha}_0}}{\alpha_0-{\hat \alpha}_0}\alpha_1
\\0& e^{{\hat \alpha}_0} \end{array}\right]~~.
\ee
The group $G$ contains an Abelian subgroup $T\approx \C^*\times \C^*$ 
which consists of elements of the form:
\be
g=\left[\begin{array}{cc}
g_0& 0\\0& {\hat g}^0 \end{array}\right]=
\left[\begin{array}{cc}
e^{\alpha_0} & 0
\\0& e^{{\hat \alpha}_0} \end{array}\right]~~.
\ee
On the other hand, elements of the type:
\be
g=\left[\begin{array}{cc}
1& g_1\\0& 1\end{array}\right]=\left[\begin{array}{cc}
1& \alpha_1\\0& 1\end{array}\right]~~
(g_1=\alpha_1\in \Omega^1_{harm}(L)\approx \C^3)~~
\ee
form an Abelian 
normal subgroup $N$ which is isomorphic with the three-dimensional 
complex translation group $(\C^3, +)$. 
The group $G$ can be viewed as a semi-direct product between $T$ and $N$. 
Note the real form of $G$ is not compact. 

Given an element 
$u=\left[\begin{array}{cc}u_1&u_2\\
u_0&{\hat u}_1
\end{array}\right]\in K^1$, $G$ acts on it via its adjoint representation:
\be
\label{adj}
Ad_g(u)=g\bullet u \bullet g^{-1}=
\left[\begin{array}{cc}
u_1 +\frac{u_0}{g_0} g_1& ~~~~\frac{g_0}{{\hat g}_0} 
u_2+\frac{g_1\wedge (u_1-{\hat u}_1)}{g_0}\\
\frac{{\hat g}_0}{g_0} u_0& {\hat u}_1+ \frac{u_0}{g_0}g_1
\end{array}\right]~~.
\ee

\subsubsection{The potential and its critical variety} 

In the case under consideration, the matrices $X_i, Y_i, Z_{ij}$ and $W$ in 
equation (\ref{utorus}) reduce to complex constants:
\be
X_i:=\xi_i~~,~~Y_i:=\iota_i~~,~~Z_{ij}:=\zeta_{ij}~~,~~W=\omega~~.
\ee
The holomorphic potential (\ref{Wtorus}) becomes:
\bea
W(u)=
(2\pi)^3\omega\left[\zeta_{12}(\xi_3-\iota_3)+\zeta_{23}(\xi_1-\iota_1)+
\zeta_{31}(\xi_2-\iota_2)\right]~~,
\label{holopot1wrap}
\eea
while equations (\ref{Z}) give:
\bea
\label{Z_s}
&&u_0u_2=0\Leftrightarrow 
\zeta_{12}\omega=\zeta_{23}\omega=\zeta_{31}\omega=0~~\nn\\
&&u_0(u_1-{\hat u}_1)=0\Leftrightarrow
\omega(\xi_1-\iota_1)=\omega(\xi_2-\iota_2)=\omega(\xi_3-\iota_3)=0~~\\
&& u_2\wedge (u_1-{\hat u}_1)=0\Leftrightarrow 
\zeta_{12}(\xi_3-\iota_3)+
\zeta_{23}(\xi_1-\iota_1)+\zeta_{31}(\xi_2-\iota_2)=0~~.\nn
\eea
The variety ${\cal Z}\subset \C^{10}(\xi_i,\iota_i, \zeta_{ij}, \omega)$ 
defined by these conditions has two 
irreducible components ${\cal Z}_1$ and ${\cal Z}_2$ 
described by the equations:
\be
{\cal Z}_1:~~u_0=0\Leftrightarrow\omega=0~~,~~
u_2\wedge (u_1-{\hat u}_1)=0\Leftrightarrow 
\zeta_{12}(\xi_3-\iota_3)+\zeta_{23}
(\xi_1-\iota_1)+\zeta_{31}(\xi_2-\iota_2)=0~~.\nn
\ee
and:
\be
{\cal Z}_2:~~u_2=0\Leftrightarrow \zeta_{ij}=0~~,~~
u_1={\hat u}_1\Leftrightarrow \xi_i=\iota_i~~\nn
\ee
It is clear that ${\cal Z}_1$ and ${\cal Z}_2$ have complex 
dimension $8$ and $4$ respectively. These varieties intersect along
the 3-dimensional locus:
\be
{\cal Z}_3 ={\cal Z}_1\cap {\cal Z}_2:u_0=u_2=0~~,~~u_1={\hat u}_1~~,
\ee
which is parameterized by $\xi_i$. 
In fact, ${\cal Z}_2$ is a copy of $\C^4$, while ${\cal Z}_1$ is a 
singular quadric which can be viewed as a fibration over $\C^6$
via the map  $\pi : (u_1, {\hat u}_1, u_2)\rightarrow (u_1, {\hat u}_1)$. 
The generic fiber is a copy of $\C^2$, described by the equation 
$u_2\wedge (u_1-{\hat u}_1)=0$ for $u_1-{\hat u}_1\neq 0$. 
Upon defining $q_i:=\xi_i-\iota_i$ and\footnote{
We let $\zeta_{ji}:=\zeta_{ij}$ for $j>i$, so that $\zeta_1=\zeta_{12}$
etc. Then $\zeta=\zeta_1 dx^2\wedge dx^3+\zeta_2 dx^3\wedge dx^1+
\zeta_3 dx^1\wedge dx^2$.} 
$\zeta_i:=\frac{1}{2}\epsilon_{ijk}\zeta_{jk}$, this condition becomes:
\be
\label{zetaq}
\boldzeta\cdot {\bf q}=0~~,
\ee
where $\boldzeta$ and ${\bf q}$ are complex vectors of 
components $\zeta_i$ and $q_i$, and $\cdot $ is the 
natural complex-bilinear product in $\C^3$
(namely ${\bf q}\cdot {\bf b}=a_ib_i$). 
The $\C^2$ fiber degenerates above the 
discriminant locus $\Delta\subset \C^6$ defined by 
the equations ${\bf q}=0\Leftrightarrow u_1={\hat u}_1$
(the diagonal in the product 
$\C^6=\C^3(u_1)\times \C^3({\hat u}_1)$), where 
it becomes a copy of $\C^3$. 
The variety ${\cal Z}_1$ is singular 
along the zero section of the resulting 
$\C^3$-bundle, which coincides with the intersection 
${\cal Z}_3={\cal Z}_1  \cap {\cal Z}_2$. 
This intersection is a copy of $\C^3$, sitting above $\Delta$.

\subsubsection{The moduli space}

Using (\ref{adj}), it is easy to check that the adjoint action
of $G$ 
preserves each of the components ${\cal Z}_1$ and ${\cal Z}_2$, on which it
reduces to the forms:
\bea
&&Ad_g\left[\begin{array}{cc}
u_1 &u_2\\
0 & {\hat u}_1
\end{array}\right]
=\left[\begin{array}{cc}
u_1 & ~~~\frac{g_0}{{\hat g}_0} u_2+\frac{g_1\wedge (u_1-{\hat u}_1)}{g_0}\\
0 & {\hat u}_1
\end{array}\right]~~,{\rm~on~}~~{\cal Z}_1~~,\\
&&Ad_g\left[\begin{array}{cc}
u_1&  0\\
u_0& u_1
\end{array}\right]=
\left[\begin{array}{cc}
u_1 +\frac{u_0}{g_0} g_1&  0\\
\frac{{\hat g}_0}{g_0} u_0& u_1+ \frac{u_0}{g_0}g_1
\end{array}\right]~~~~,{\rm~on~}~~{\cal Z}_2~~,\nn\\
&&Ad_g =id ~~,{\rm~on~}~~{\cal Z}_3 ~~.\nn
\eea
The diagonal subgroup 
$T_d:=\{\left[\begin{array}{cc}g_0 &0\\0&g_0\end{array}\right]
|g_0\in \C^*\}\subset T$ acts trivially on ${\cal Z}_2$, while the 
action of $G/T_d$ is transitive and fixed-point free 
on ${\cal Z}_2-{\cal Z}_3$. 
It follows that the quotient ${\cal M}_0$ 
of ${\cal Z}_2-{\cal Z}_3$ by $G/T_d$ is simply 
a point, which we denote by $p$:
\be
{\cal M}_0=\{p\}~~.
\ee  

It is also easy to see that $G$ acts only along the fibers of 
the fibration 
${\cal Z}_1\stackrel{\pi}{\rightarrow}\C^3(u_1)\times \C^3({\hat u}_1)$.
To understand this action, let $g_1=G_i dx^i$ and consider the 
complex vector ${\bf g}:=(G_1, G_2, G_3)$. Using the notations in 
equation (\ref{zetaq}), the group action on a fiber which does not sit 
above $\Delta$ leaves $u_1$ and ${\hat u}_1$ unchanged, and modifies 
$\boldzeta$ (and thus $u_2$) according to Figure 5:
\be
\boldzeta\rightarrow \boldzeta' =\frac{g_0}{{\hat g}_0}\boldzeta+
\frac{1}{g_0}{\bf g}\times {\bf q}~~.
\ee

\begin{figure}[hbtp]
\begin{center}
\scalebox{0.5}{\input{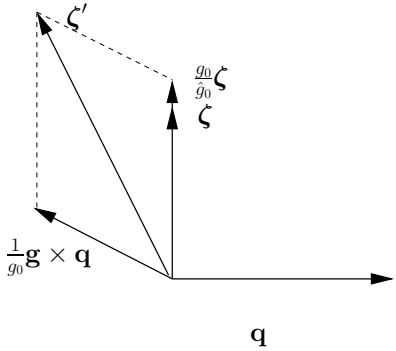}}
\end{center}
\caption{Action of $G$ along the generic (i.e. $\C^2$) 
fibers of ${\cal Z}_1$. The vector 
${\bf g}=(G_1,G_2,G_3)$ is defined by the components $G_i$ of the one-form
$g_1=G_idx^i$.}
\end{figure}

The two-dimensional subgroup $G_1$ consisting of elements of the form:
\be
g=\left[\begin{array}{cc} g_0&~~\alpha q_1\\0&g_0\end{array}\right]~~,
g_0\in \C^*, \alpha\in \C
\ee
acts trivially on the $\C^2$-fibers, while $G/G_1$ acts 
transitively\footnote{Every $\boldzeta$ is 
stabilized by elements 
of the form 
$g=\left[\begin{array}{cc} g_0&\frac{g_0(g_0^2-1)}{{\bf q}^2}
\\0&g^{-1}_0\end{array}\right]$ (with $g_0^2\neq 1$), 
which form a one-dimensional subgroup of $G/G_1$ .}.
Thus the quotient of ${\cal Z}_1-\pi^{-1}(\Delta )$ by $G/G_1$ is 
(topologically) a copy of $\C^3(u_1)\times \C^3({\hat u}_1)-\Delta$. 
On the $\C^3$ fibers, the action of $G$ 
reduces to the standard  $\C^*$ action by homotheties (the 
rescaling $u_2\rightarrow \lambda u_2, \lambda\in \C^*$).
Hence the quotient of $\pi^{-1}(\Delta)-{\cal Z}_3$ 
is a $\P^2$-bundle above $\Delta$.

Finally, we have ${\cal Z}_3\approx \Delta \approx 
\C^3$, on which $G$ acts trivially. This simply gives a copy of $\Delta$.
Putting these pieces together, we obtain a copy ${\cal M}_1$ of 
$\C^6=\C^3(u_1)\times \C^3({\hat u}_1)$, and a $\P^2$-fibration over 
$\Delta$ which we denote by ${\cal M}_2$. 

Summarizing, the 
moduli space ${\cal M}$ consists of three components: 
${\cal M}_0=\{p\}$, ${\cal M}_1=\C^3(u_1)\times 
\C^3({\hat u}_1)$ 
and ${\cal M}_2$, which is a $\P^2$-fibration over the diagonal $\Delta$ of ${\cal M}_1$.
By the discussion above, configurations in ${\cal M}_1$ admit representatives
of the form $u=\left[\begin{array}{cc} u_1&0\\0&{\hat u}_1\end{array}\right]$,
and thus correspond (up to a gauge transformation) to deformations of the
diagonal components of the original background; this generates 
the moduli space of flat connections $A_a$ and $A_b$ on the two 
original D-branes, and corresponds to deforming them independently without 
condensing boundary condition changing fields.
Points of ${\cal M}_2$ admit representatives of the form
 $u=\left[\begin{array}{cc} u_1&u_2\\0&u_1\end{array}\right]$ (with 
$u_2$ determined up to a rescaling), 
and correspond to moduli obtained by condensation of the 
two-form, starting from 
a diagonal background with equal connections $A_a=A_b$.
The effect of turning on $u_2$ is to blow up the diagonal 
$\Delta$ in the product ${\cal M}_1=\C^3(u_1)\times \C^3({\hat u}_1)$
(indeed, $({\cal M}_1-\Delta)\cup {\cal M}_2$ coincides with the blow-up 
of ${\cal M}_1$ along its diagonal). Finally, configurations 
associated with ${\cal M}_0$ can be gauge-transformed to the form
 $u=\left[\begin{array}{cc} 0&0\\1&0\end{array}\right]$, and 
correspond to turning on $u_0$. As discussed above, this produces 
an acyclic composite of the two D-branes, thereby leading to the 
isolated point $p$ of the moduli space.  This situation is described in 
Figure 6.

\begin{figure}[hbtp]
\begin{center}
\scalebox{0.5}{\input{moduli.pstex_t}}
\end{center}
\caption{The local structure of the moduli space.}
\end{figure}

\subsection{Fine structure of the moduli space for unit defect}

Consider the case $r_a=r_b=n+1$ and $d=1$. 
In this situation, we can choose bases for the fibers of $E_a$ and $E_b$
such that $\phi_0$ has the form:
\be
\phi_0=\left[\begin{array}{cc}0&0^T_n\\0_n &\id_{n\times n}
\end{array}\right]~~,
\ee
where $0_n$ is a column vector with vanishing entries and $0^T_n$ is 
its transpose. In this case $K(\phi^0)$ and $I^\perp (\phi_0)$ are 
both represented by vectors of the form $\left[\begin{array}{c}\xi\\0_n
\end{array}\right]$. Then the matrices $X_i, Y_j, Z_{ij}$ and $W$
have the forms:
\be
X_i=\left[\begin{array}{cc}\xi_i&0_n^T\\
0_n&0_{n\times n}\end{array}\right]~~,~~
Y_i=\left[\begin{array}{cc}\iota_i&0_n^T\\
0_n&0_{n\times n}\end{array}\right]~~,~~
Z_{ij}=\left[\begin{array}{cc}\zeta_{ij}&0_n^T\\
0_n&0_{n\times n}\end{array}\right]~~,~~
W=\left[\begin{array}{cc}\omega&0_n^T\\
0_n&0_{n\times n}\end{array}\right]~~,
\ee
with $\xi_i, \iota_i, \zeta_{ij}$ and $\omega$ some complex constants.
Inserting them in equation (\ref{Wtorus}) or directly 
using (\ref{W_scalar}) we find that this choice of $\phi_0$ can be described
equally well using the matrix-valued form
\be
u=\left[\begin{array}{cc} u_1&u_2 \\u_0& {\hat u}_1\end{array}\right]
\in 
[\Omega^*(L,End(K(\phi_0)\oplus I^\perp(\phi_0))]^1~~,
\ee
with:
\bea
&&u_1=\xi_idx^i~~,~~~{\hat u}_1=\iota_i dx^i~~,~~~u_0=\omega~~\nn\\
&&u_2=\zeta_{12}dx^1\wedge dx^2+\zeta_{23}dx^2\wedge dx^3+
\zeta_{31}dx^3\wedge dx^1~~.
\eea
This observation reduces the analysis to the case considered in the 
previous section. Indeed, the holomorphic potential takes the form
(\ref{holopot1wrap}) and consequently, the stratum corresponding to $d=1$ 
in the moduli space of graded D-branes with unit relative grading and 
equal rank bundles, wrapping special Lagrangian tori has the structure 
described in Figure 6. Turning on $u_0=\omega$ leads to formation
of acyclic composites and disappearance of open string states form the theory.
This is not unexpected since by turning on $u_0$ corresponds to a shift the 
background to the case of $d=0$ and, as we showed in 
section \ref{acyclic_general}, this leads to acyclic composites.

\section{Conclusions}

We discussed deformations of systems of graded topological D-branes.
Upon considering the associated string field theory, we constructed 
a physically motivated quantity which generalizes the holomorphic 
potential of \cite{Douglas_quintic, Kachru, superpot} and showed 
that it leads to an equivalent description of the deformation problem, 
which is often more efficient from a computational point of view. 
Upon applying these methods to topological D-brane pairs of unit relative 
grade, we gave a general proof that scalar condensation in such systems 
leads to acyclic composites in the case when the underlying flat connections 
are equivalent. For the case of singly-wrapped branes on 3-tori, we gave 
an explicit local construction of the moduli space, confirming the 
existence of a 
branch parameterized by a two-form. This shows that such condensation processes
are not (completely) obstructed at the topological level, and underscores 
the need for a deeper understanding of their role in the construction 
of topological D-brane categories and  their subcategories of stable D-branes.
While our discussion has been limited to the large radius limit of the 
A-model\footnote{Our reason for considering the A-model is that 
the underlying geometry is {\em more} complicated in this case--due to the
nontrivial topology of special Lagrangian 3-cycles. The B-model is 
conceptually simpler.}, it is clear that 
a similar analysis goes through for the B-model case. In that situation, one 
obtains a holomorphic potential which allows for an explicit 
description of deformation problems in the derived category. 

The study of deformations of D-brane composites is of crucial importance 
for the program of \cite{Douglas_Kontsevich} and for gaining a better 
understanding of the extended moduli space of open strings. 
The point of view adopted in this paper follows the approach of 
\cite{com1, com3, sc} by retreating to the underlying string field theory, 
which allows for a standard description of deformations in 
terms of Maurer-Cartan equations. For both the A and B models, it is possible 
to pass from this description to one in terms of triangulated categories, 
upon dividing through quasi-isomorphisms (this amounts to keeping only 
the data which is invariant under infinitesimal canonical 
transformations in the 
BV formalism, and is in many ways only a `local' description). 
The latter point of view does not seem to 
allow for a direct formulation 
of the deformation problem. For example, it is not immediately clear how one 
can define deformations of an object in the derived category of coherent 
sheaves\footnote{It is of course trivial to define virtual, or infinitesimal, 
deformations by considering $Ext$ groups. However, one expects such 
deformations to be obstructed, and it is not apriori clear how to describe
the relevant obstructions -- knowledge of virtual deformations 
tells us little about the true moduli space of an object. 
In our approach, the effect of obstructions is described by the 
potential $W$, which carries over to the derived category. It is only through 
this remnant of the original, string field description, that one knows how 
to describe true deformations at the derived category level.} 
(the relevant triangulated category for the B-model), 
which is proposed as a description of B-model topological D-brane 
physics. One of the major virtues of the string field theory approach is that 
it allows for an entirely natural formulation of deformations, in a language 
which is both physically and mathematically well-established. 
This description can be carried over to the derived category, {\em provided} 
that one endows the latter with the extra datum induced by the  $A_\infty$ 
products of Section 4. In fact, it seems that any  description 
of the deformation problem at the derived category level 
must consider such supplementary input. 
This vindicates the point of view (advocated in \cite{BK}) that the 
correct object of study is an {\em enhanced} version of the derived category, 
which `remembers' the string field theoretic data. One could as 
well study the topological string field theory itself.

\acknowledgments{We wish to thank D.~Vaman for collaboration in the initial 
stages of this project. 
We are indebted to M.~Rocek
for support and interest in our work.  C.I.L. thanks Rutgers University 
(where part of this paper was completed) 
for hospitality and providing excellent conditions. He also thanks M. Douglas
for an enlightening conversation.
R.R. thanks the C.N. Yang 
Institute for Theoretical Physics for support during 
the initial stages of this project.
The present work was supported by the Research Foundation under NSF 
grants PHY-9722101, NSFPHY00-98395 (6T) and DOE grant DOE91ER40618 (3N).}

\

\appendix
\section{The deformed Laplacian for 
a D-brane pair in a scalar background}

Consider a graded D-brane pair (of unit relative grade) in a scalar 
background, as in section \ref{example}. As in Section 3, 
we assume that the background flat connections on the bundles $E_a$ and 
$E_b$ are unitary with respect to the auxiliary metrics carried by these 
bundles. With these assumptions, we prove 
the relation:
\be
\label{Delta_app}
\Delta_\phi u=\Delta u+\{[\phi,\phi^\dagger]_\bullet ,u\}_\bullet ~~,
\ee
which was used in Section \ref{deformedlap}.
We shall prove (\ref{Delta_app}) in a three steps:

\

(1) First, we notice that:
\be
\label{ccom}
c[\phi,u]_\bullet=\{cu,\phi^\dagger\}_\bullet~~,~~{\rm~for~all}~~
u\in {\cal H}~~.
\ee
This relation follows directly from the definition of $c$ and $\bullet$ and 
relations such as:
\bea
~*(\phi\bullet u)^\dagger&=&(-1)^{rk u}*(\phi\circ u)^\dagger=
 (-1)^{rk u} (*u^\dagger)\circ \phi^\dagger=(-1)^{rk u}(*u^\dagger)\bullet 
\phi^\dagger~~\nn\\
~*(u\bullet \phi)^\dagger&=&*(u\circ \phi)^\dagger=\phi^\dagger 
\circ (*u^\dagger)=
(-1)^{rk u +1}\phi^\dagger \bullet (*u^\dagger)~~,
\eea
which make use of the fact that the only non-vanishing component of $\phi$ 
is a zero-form. 
\

(2) Using (\ref{ccom}), we compute:
\be
\label{dphidag}
d_\phi^\dagger u=(-1)^{|u|}cd_\phi c u=d^\dagger u+(-1)^{|u|}c[\phi,cu]_\bullet =
d^\dagger u+\{\phi^\dagger,u\}_\bullet ~~.
\ee 
In particular, the adjoint of the operator $A_\phi u=[\phi,u]_\bullet$ 
is $A_\phi^\dagger u=\{\phi^\dagger, u\}_\bullet$. 

\

(3) We have:
\bea
\label{dd}
d^\dagger_\phi d_\phi u &=& d^\dagger d u +d^\dagger [\phi, u]_\bullet+
\{\phi^\dagger, du\}_\bullet +\{\phi^\dagger, [\phi,u]_\bullet \}_\bullet~~\nn\\
d_\phi d^\dagger_\phi u &=& dd^\dagger u +[\phi, d^\dagger u]_\bullet+
d\{\phi^\dagger, u\}_\bullet +[\phi, \{\phi^\dagger,u\}_\bullet ]_\bullet~~,
\eea
and:
\bea
d^\dagger[\phi,u]_\bullet=-[\phi,d^\dagger u]_\bullet ~~,~~d\{\phi^\dagger, u\}_\bullet =
-\{\phi^\dagger, du\}_\bullet~~
\eea
(these relations use the assumption that $A_a$ and $A_b$ are unitary 
connections).

Hence adding the two equations in (\ref{dd}) gives:
\be
\Delta_\phi u=\Delta u+\{\phi^\dagger, [\phi,u]_\bullet \}_\bullet +
[\phi, \{\phi^\dagger,u\}_\bullet ]_\bullet =
\Delta u +\{[\phi, \phi^\dagger]_\bullet , u\}_\bullet ~~.
\ee
This establishes (\ref{Delta}). 

\section{Another approach to the tree level potential}

This appendix  gives an alternate derivation of the potential of Section 4.  
This is a textbook exercise \cite{siegel}, but it is instructive 
to see how the potential arises from standard field theory techniques. 
There are two equivalent methods for constructing the tree-level potential: 
the Feynman diagram expansion of Section 4 and the JWKB  approximation.
Here we describe the second approach. 

The gauge fixing condition (\ref{gc}) implies that the 
field $\phi$ belongs to the subspace
$(im d_A^\dagger\oplus K_A)\cap {\cal H}^1$.
Since we wish to integrate out modes belonging to $(im\,d_A^\dagger)^1$, we split
\footnote{We emphasize that, because of 
our choice of background, this is {\em not} the standard 
splitting one uses in the background field formalism.} $\phi$ 
into a ``background part'' $\phi_K\in K_A^1$ and a ``quantum part'' 
$\psi\in (im\,d_A^\dagger)^1$. 
With this decomposition, 
the classical action (\ref{action}) takes the form:
\be
S(\phi)=S(\phi_K+\psi)=S_0(\psi)+gS_I(\phi_K+\psi)
\ee
where $S_0, S_I$ are the quadratic and cubic terms.
We have artificially introduced an adiabatic parameter 
$g$. We will set  $g=1$ at the end.

The partition function computes the potential for $\phi_K$:
\be
Z[\phi_K] = \int {\cal D}\psi e^{-{i\ov\hbar}(
S_0(\psi)+g S_I(\phi_K+\psi))} =e^{-{i\ov\hbar}gW_{full}(\phi_K)}~~.
\label{partition}
\ee
At this stage one can use the saddle point approximation 
to express the tree-level part $W$ of $W_{full}$ as the classical action $S$ evaluated on a solution 
$\phi_{cl}(\phi_K)=\phi_K+\psi_{cl}(\phi_K)$ to the classical equation of motion 
$d\phi_{cl}+g\phi_{cl}\bullet \phi_{cl}=0$ which has the property that $\psi_{cl}$ vanishes 
in the adiabatic limit $g=0$.
Given such a solution, one has $\langle \phi_{cl}, \phi_{cl}\bullet \phi_{cl}\rangle=
-\langle \phi_{cl}, d\phi_{cl}\rangle$ (for $g=1$), so that:
\be
\label{part}
W(\phi_K)=S(\phi_{cl})=\frac{1}{6}\langle \phi_{cl}, d\phi_{cl}\rangle=\frac{1}{6}\langle 
\psi_{cl}, d\psi_{cl}\rangle~~, 
\label{ugly}
\ee
where we used invariance of the bilinear form $\langle .,. \rangle$ with respect to 
the differential $d$ and the fact that $d\phi_K=0$. This relation is somewhat inconvenient for 
the computation of $W$, since its adiabatic expansion 
will involve a double sum.

To avoid this problem, one can proceed by constructing the 
quantity:
 \be
-{i\ov \hbar}{\delta\ov\delta\phi_K}W_{full}(\phi_K)={1\ov gZ[\phi_K]}
{\delta\ov\delta\phi_K}Z[\phi_K]
\ee
where we use the convention that the functional derivative of a functional 
$F[\phi]$ is defined through:
\be
\delta F[\phi]=\langle \delta\phi, \frac{\delta F}{\delta \phi}\rangle 
=\int_{L}{str\left[\delta\phi(x)\bullet \frac{\delta F}{\delta \phi(x)}\right]}
~~.
\ee

Equation (\ref{fullpot}) gives:
\begin{eqnarray} 
{\delta\ov\delta\phi_K}W_{full}(\phi_K)=P \bigg(
\phi_K\bullet\phi_K+\langle\psi\rangle_{\phi_K}\bullet\phi_K+
\phi_K\bullet\langle\psi\rangle_{\phi_K}+\langle\psi\bullet\psi
\rangle_{\phi_K}\bigg)
\label{ed}
\end{eqnarray} 
where the expectation value of 
a functional $f[\psi(x)]$ in the background $\phi_K$ is 
defined through:
\be
\langle f[\psi(x)]\rangle_{\phi_K}={1\over Z[\phi_K]}   
\int {\cal D}\psi~ f[\psi(x)] ~e^{-{i\ov\hbar}(S_0(\psi)+
g S_I(\phi_K+\psi))}~~. 
\ee
and $P$  is the orthogonal projector on $K$, as introduced in Section 4.3.

Equation (\ref{ed}) is valid to all loop orders.  To isolate the tree level part $W$, we
use the saddle point approximation which tells us that the tree-level 
contribution to $\langle\psi(x)\rangle_{\phi_K}$ and 
$\langle\psi(x)\bullet \psi(x)\rangle_{\phi_K}$ 
is given by the solution $\psi_{cl}$ to the classical equation of motion which vanishes
in the limit of vanishing $g$. It is much easier to compute $W$ using 
this approach than from equation (\ref{part}). 

To solve the classical equation of motion, 
we write $\psi_{cl}$ as a formal power series in $g$:
\be
\psi_{cl}=\sum_{n\geq 2} g^{n-1}\psi_{n}(\phi_K^{\otimes n})~~.
\label{texp}
\ee
This is a Taylor expansion, with the adiabatic parameter included explicitly. 
Since $\psi_{cl}$ 
is constrained to belong to $im d^\dagger$, so are all of its 
coefficients $\psi_n(\phi_K^{\otimes n})$.
Using expansion (\ref{texp}), the first 
derivative of the tree-level potential takes the form:
\bea
\label{potexp}
{\delta W\ov\delta\phi_K}&=&P\bigg(\phi_K\bullet\phi_K+g(\psi_2\bullet\phi_K+
\phi_K\bullet\psi_2)\\
&+&\sum_{n\ge 3}g^{n-1}\Big(\psi_n\bullet\phi_K+
\phi_K\bullet\psi_n+\sum_{p+q=n+1}\psi_p\bullet\psi_q\Big)\bigg)~~.\nn
\eea

Upon substituting (\ref{texp}) in the classical field 
equation $d\psi_{cl}+g(\phi_K+\psi_{cl})\bullet (\phi_K+\psi_{cl})=0
\Leftrightarrow d\psi_{cl} =
-g\pi_d[(\psi_K +\psi_{cl})\bullet (\phi_K+\psi_{cl})]$ (recall  that $\pi_d$ is 
the projector on $im d$) and matching powers of $g$, we find a recurrence 
relation for $\psi_n:=\psi_n(\phi_K^{\otimes n})$:
\bea
d\psi_2&=&-\pi_d(\phi_K\bullet \phi_K)~~\nn\\
d\psi_n&=& -\pi_d\left(\phi_K \bullet \psi_{n-1} + \psi_{n-1}\bullet \phi_K 
                        +\sum_{l + m = n} \psi_l\bullet \psi_m \right)~~{\rm for}~~n\geq 3 ~~.
\eea
Since $\psi_n$ belong to $im d^\dagger$,  
and the restriction $d:im d^\dagger\rightarrow im d$ is 
invertible, these relations can be solved as:
\bea
\psi_2&=&-\frac{1}{d}\pi_d(\phi_K\bullet \phi_K)=-U(\phi_K\bullet \phi_K)~~\nn\\
\psi_n&=& -U\left(\phi_K \bullet \psi_{n-1} + \psi_{n-1}\bullet \phi_K 
                        +\sum_{l + m = n} \psi_l\bullet \psi_m \right)~~{\rm for}~~n\geq 3 ~~.
\label{recrel}
\eea

As shown in figure 7, this 
equation can be represented through a sum of tree-level graphs (for convenience, we define 
$\psi_1:=\phi_K$). 
\begin{figure}[hbtp]
\begin{center}
\mbox{\epsfxsize=9truecm \epsffile{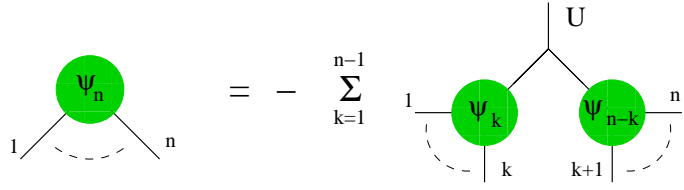}}
\end{center}
\caption{Tree graph expansion of the perturbative solution to the 
classical field equation.\label{trees}}
\end{figure}
This graphical representation makes it intuitively clear that the object
$\langle \phi_K, (\psi_{n}\bullet\phi_K+\phi_K\bullet\psi_{n}+
\sum_{p+q=n+1}\psi_p\bullet\psi_q)\rangle$ is the diagonal value of 
a cyclically symmetric multilinear form, since it is a 
`complete' sum of tree graphs with $n+1$ external lines. 
It is possible to justify this claim by direct computation.
Instead, we show cyclicity by relating $\psi_n$ to the products 
$\lambda_n$ used in reference \cite{superpot}. 
To this end, we write:
\be
\psi_n(\phi_K^{\otimes n})=(-1)^{n(n+1)/2}U\lambda_n(\phi_K^{\otimes n})~~,
\label{redef}
\ee
which brings (\ref{recrel}) to the form:
\bea
\label{recl}
\lambda_2&=&\phi_K\bullet \phi_K~~\\
{\lambda}_n &=& 
(-1)^{n-1} \phi_K \bullet U{\lambda}_{n-1} +  (-1)^{n-1}
U{\lambda}_{n-1}\bullet \phi_K 
                        -\sum_{l + m = n} (-1)^{m l}U{\lambda}_l\bullet U{\lambda}_m ~~{\rm for}~~
n\geq 3 ~~,\nn
\eea
where we denoted $\lambda_n(\phi_K^{\otimes n})$ by $\lambda_n$.
This coincides with the recurrence relation of \cite{superpot}, 
written for the particular case of fields with unit $U(1)$ charge. 
This redefinition also brings the diagrams of Figure 7 to the Feynman form,  since we now 
have a propagator for each internal line. 
Equation (\ref{potexp}) gives (for $g=1$): 
\be
\frac{\delta W(\phi_K)}{\delta\phi_K}
=-\sum_{n\ge 2}(-1)^{{1\ov 2}n(n+1)}P\lambda_n(\phi_K^{\otimes n})~~.
\ee

Using the cyclicity properties of $\lambda_n$ discussed in \cite{superpot}
we find:
\be
W(\phi_K)=-\sum_{n\ge 2}{\frac{(-1)^{n( n+1)/2}}{n+1}\langle\phi_K,\,
P\lambda_n(\phi_K^{\otimes n})\rangle}=
-\sum_{n\ge 3}{\frac{(-1)^{n( n-1)/2}}{n}\langle\phi_K,\,
r_{n-1}(\phi_K^{\otimes {n-1}})\rangle}~~.
\ee

This concludes the derivation of the potential.

\end{document}